\documentclass[3p]{elsarticle}
\usepackage{lineno,hyperref}
\usepackage{graphicx}
\usepackage{amsmath}
\usepackage{color}
\usepackage{mathtools}
\usepackage[normalem]{ulem}
\usepackage{rotating}
\usepackage{makecell}
\usepackage{threeparttable, tablefootnote}
\newcommand{\Beta}{\mathrm{B}}

\journal{arXiv}
\begin{document}

\begin{frontmatter}

\title{Implied and Realized Volatility: A Study of Distributions and the Distribution of Difference}

 \author[mymainaddress]{M. Dashti Moghaddam}
 \author[mymainaddress]{Jiong Liu}
 \author[mymainaddress]{R. A. Serota\fnref{myfootnote}}
 \fntext[myfootnote]{serota@ucmail.uc.edu}
 
 \address[mymainaddress]{Department of Physics, University of Cincinnati, 
 Cincinnati, Ohio 45221-0011}

\begin{abstract}
We study distributions of realized variance (squared realized volatility) and squared implied volatility, as represented by VIX and VXO indices. We find that Generalized Beta distribution provide the best fits. These fits are much more accurate for realized variance than for squared VIX and VXO -- possibly another indicator that the latter have deficiencies in predicting the former. We also show that there are noticeable differences between the distributions of the 1970-2017 realized variance and its 1990-2017 portion, for which VIX and VXO became available. This may be indicative of a feedback effect that implied volatility has on realized volatility. We also discuss the distribution of the difference between squared implied volatility and realized variance and show that, at the basic level, it is consistent with Pearson's correlations obtained from linear regression.
\end{abstract}


\end{frontmatter}

\section{Introduction}

\begin{figure}[!htbp]
\centering
\includegraphics[width = .49 \textwidth]{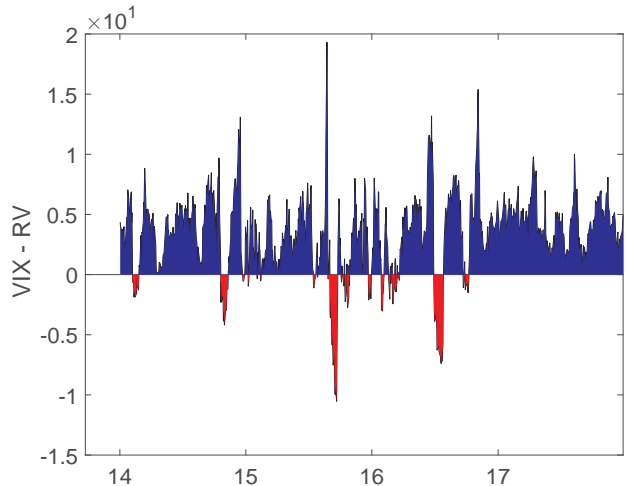}
\caption{$ {VIX - RV} $, from Jan 1st, 2014 to Dec 29th, 2017}
\label{VIX-RVPlot}
\end{figure}

Since the original volatility index (presently VXO) was introduced in 1990 by CBOE and then reintroduced in 2003 (presently VIX) the question of how well these indices predict future realized volatility (RV) remains of interest to researchers \cite{christensen1998relation, vodenska2013understanding, kownatzki2016howgood, russon2017nonlinear}. In previous two articles \cite{dashti2018implied, dashti2018ratio} we visually compared the probability density function of realized variance ($RV^2$) -- squared RV -- and squared implied volatility, as represented by VIX and VXO. We also studied the distribution of the ratio of $RV^2$ to $VIX^2$ and to $VXO^2$, which provided additional insights relative to qualitative comparison and simple regression analysis. Here we address specifically the form of these distributions. We also investigate the distributions of $VIX^2-RV^2$ and $VXO^2-RV^2$ due to recent interest in looking at the time series of $VIX-RV$ -- see Fig. \ref{VIX-RVPlot} -- which is equivalent to the one shown in the Wall Street Journal \cite{sindreu2018main}. 

Realized variance (index) is defined as follows
\begin{equation}
RV^2=100^2\times\frac{252}{n}\sum_{i=1}^nr_i^2
\label{RV2}
\end{equation}
where
\begin{equation}
r_i=\ln\frac{S_{i}}{S_{i-1}}
\label{ri}
\end{equation}
are daily returns and $S_{i}$ is the reference (closing) price on day $i$. This is an annualized value, where $252$ represents the number of trading days. Specifically for monthly returns $n\approx21$, $365/252\approx30/21\approx1.4$. Since VIX and VXO are evaluated daily to forecast RV for the following month and are annualized to $365$, to properly compare the distributions of $RV^2$ to $VIX^2$ and to $VXO^2$, one should rescale the distribution of $RV^2$ with the ratio of the mean of  $VIX^2$ and $VXO^2$ to that of $RV^2$ \cite{dashti2018implied}, which is usually close to $1.4$. \footnote{Accordingly, in a more meaningful version of Fig. \ref{VIX-RVPlot} RV would be rescaled with $\sqrt{365/252}$.}

\begin{figure}[!htbp]
\centering
\begin{tabular}{cc}
\includegraphics[width = 0.49 \textwidth]{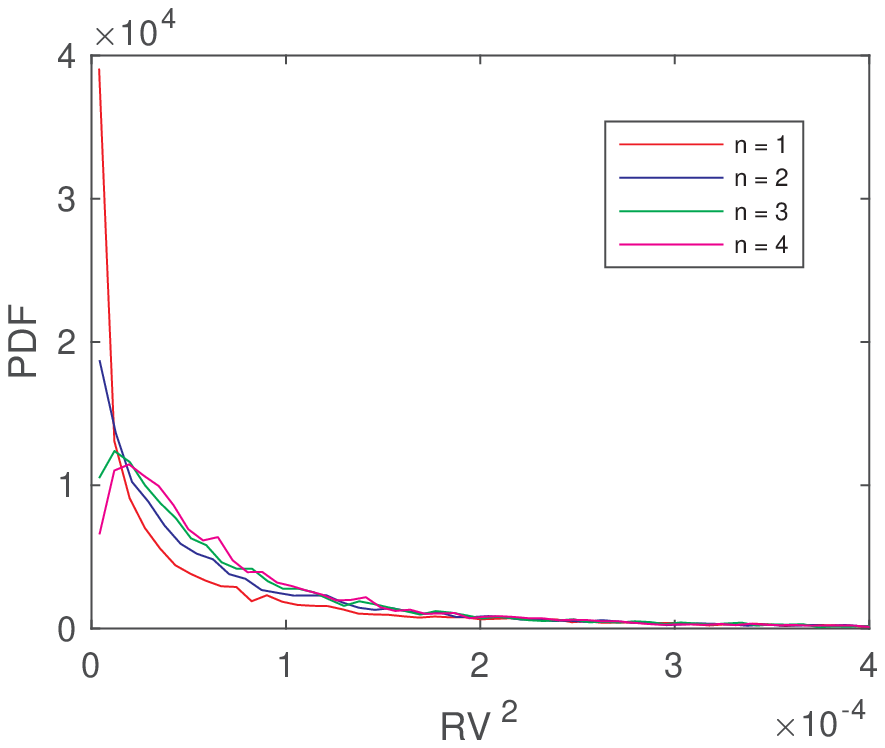} \hspace{0.1cm}
\includegraphics[width = 0.49 \textwidth]{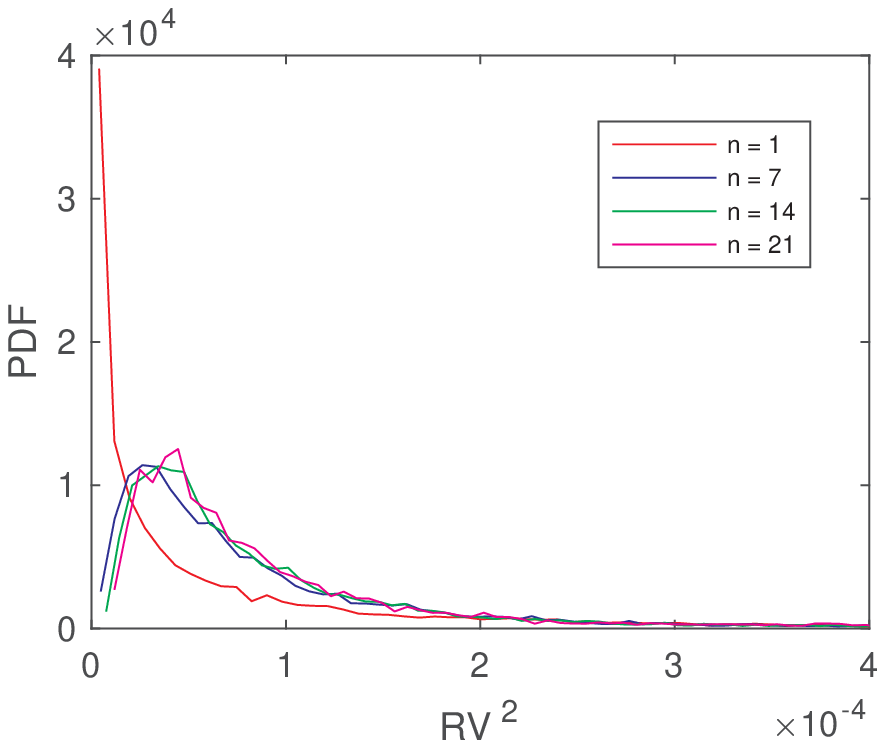}
\end{tabular}
\caption{PDFs of $\frac{1}{n}\sum_{i=1}^nr_i^2$ for $n=$1,2,3,4 (left) and $n=$1,7,14,21 (right).}
\label{sumri2}
\end{figure}

Since $RV^2$ is based on the sum of realized daily variances, the obvious questions for understanding its distribution are: what is the distribution of daily variances and what are the correlations between between them? Study of intraday returns, interpreted in terms of intraday jumps, \cite{behfar2016long} points to fat-tailed $\propto 1/x^{\mu+1}$ distributions with $1<\mu<2$. Here, our own fitting of daily realized variance $RV^2$ seems to correspond to similarly tailed distributions of returns, that is $\propto 1/x^{\frac{\mu}{2}+1}$ with $\mu$ close to the values in \cite{behfar2016long}. However, none of the distributions used here -- all based on continuous models of stochastic volatility -- are a good fit to daily $RV^2$. This is not surprising since all of continuous models are best suited for bell-shaped distributions. However, as is obvious form Fig. \ref{sumri2} it takes an addition of several days of daily $RV^2$ do develop the bell shape. Nonetheless, Generalized Beta Prime distribution (see below) provides "the best of the worst" fit to daily returns and is based on a non-mean-reverting stochastic volatility model \cite{hertzler2003classical}. 

\begin{figure}[!htbp]
\centering
\begin{tabular}{cc}
\includegraphics[width = 0.49 \textwidth]{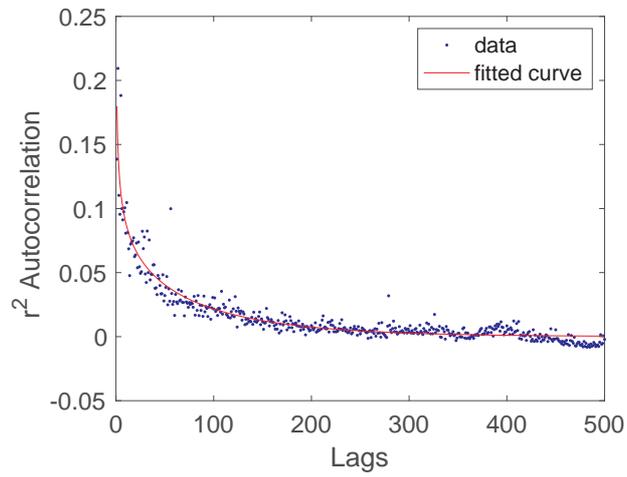}
\end{tabular}
\caption{Autocorrelation function of daily realized variance (dots) and the best fit with $c\times x^{b-1} \times exp(-a*x), a =0.0088,  b = 0.73, c = 0.18$.}
\label{returndist}
\end{figure}

Had the realized variances been uncorrelated, the monthly realized variance would have been expected, by the generalized central limit theorem, to approach a stable distribution. However, Fig. \ref{returndist} indicates otherwise, where the initial fast power-law drop-off of correlations is followed by a slow exponential decay with the time constant of about 120 days. Consequently, we are reduced to empirical fitting of the distribution function of $RV^2$  with heavy-tail distributions, including stable. We will concentrate specifically on monthly returns but Figure \ref{sumri2} shows that $RV^2$ quickly approaches its limiting form at $n \approx 5 - 7$ -- approximately the same number of days over which power law yields to exponential in Fig. \ref{returndist}.

This paper is organized as Follows. In Section \ref{distribution} we identify the list of distributions used for fitting of the probability distribution functions (PDF) of $RV^2$, $VIX^2$ and $VXO^2$ and discuss their role as steady-state distributions of stochastic differential processes. We use Maximum Likelihood Estimation  (MLE) and obtain the list of fitting parameters using the Kolmogorov-Smirnov (KS) values to compare goodness of fits. We also examine the evolution of the power-law tail exponents -- including a direct comparison of the tails -- and KS values as a function of $n$ in connection with Figs. \ref{sumri2} and \ref{returndist}. In Section \ref{difference}, we examine the distributions of differences of $VIX^2$ with scaled $RV^2$ and $VXO^2$ with scaled $RV^2$ vis-a-vis simple correlation between the indices established by linear regression. Finally, in the Appendix we look at the distributions of RV, VIX and VXO and the corresponding difference distributions, which is done because market observers and researchers are more familiar with these indices.

We use 1970-2017 S\&P 500 stock price data to calculate realized volatility and variance. Unless explicitly mentioned that we use the 1990-2017 subset of the data, the full set is used below.

\clearpage
\section {Probability Distribution Functions of $RV^2$, $VIX^2$ and $VXO^2$} \label{distribution}

As mentioned in the Introduction, we don't have analytical predictions for the distribution functions, barring the expectation that they will express fat tails. For empirical fitting we use the distributions collected in Table \ref{analyticforms}: Generalized Beta Prime (GB2), Beta Prime (BP), Generalized Inverse Gamma (GIGa), Inverse Gamma (IGa), Generalized Gamma(GGa) and Gamma (Ga). Here, $p$, $q$, $\alpha$ and $\gamma$ are shape parameters and $\beta$ is a scale parameter. We also use Stable Distribution (S)\cite{nolanstable}, $S(x; \alpha, \beta, \gamma, \delta)$, but it does not, in general, reduce to a closed-form expression. For S, $\alpha$ and $\beta$ are shape parameters, $\gamma$ is a scale parameter and $\delta$ is a location parameter. Two right columns in Table \ref{analyticforms} show the power-law exponents of the front end and of the tails respectively. GGa and Ga are included as distributions with short tails. Notice also that they are related to GIGa and IGa as distributions of the inverse variable. 

\begin{table}
\centering
\caption{Analytic Form of Probability Density Functions  for Fitting $RV^2$, $VIX^2$ and $VXO^2$}
\label{analyticforms}
\fontsize{15}{20}
\begin{tabular}{|c|c|c|c|} 
\hline
            type &       PDF&		front exponent&		tail exponent  \\
\hline
$S(x; \alpha, \beta, \gamma, \delta)$ & $ $& $ $ &$-(\alpha + 1)$\\
\hline
$GB2(x; p, q, \alpha, \beta)$ & $\frac{\alpha(1+(\frac{x}{\beta})^{\alpha})^{-p-q}(\frac{x}{\beta})^{-1+p\alpha}}{\beta \space \Beta(p,q)}$  & $\alpha p - 1$ & $-(\alpha q + 1)$\\
\hline
$BP(x; p, q, \beta)$ & $ \frac{(1+\frac{x}{\beta})^{-p-q}(\frac{x}{\beta})^{-1+p}}{\beta \space B(p,q)}$  & $p - 1$ & $-(q + 1)$\\
\hline
$GIGa(x; \alpha, \beta, \gamma)$ & $\frac{\gamma e^{-(\frac{\beta}{x})^\gamma}(\frac{\beta}{x})^{1+\alpha\gamma}}{\beta \Gamma(\alpha)}$ & $ $  &$-(\alpha\gamma + 1)$\\
\hline
$IGa(x; \alpha, \beta)$ &  $\frac{e^{-\frac{\beta}{x}}(\frac{\beta}{x})^{1+\alpha}}{\beta \Gamma(\alpha)}$&$ $ &$-(\alpha + 1)$ \\
\hline
$GGa(x; \alpha, \beta, \gamma)$ & $\frac{\gamma e^{-(\frac{x}{\beta})^{\gamma}}(\frac{x}{\beta})^{-1+\alpha \gamma}}{ \beta \Gamma(\alpha)}$ &$\alpha\gamma - 1$ &$ $ \\
\hline
$Ga(x; \alpha, \beta)$ &  $\frac{e^{-\frac{x}{\beta}}(\frac{x}{\beta})^{-1+\alpha}}{ \beta \Gamma(\alpha)}$ &$\alpha - 1$ &$ $ \\
\hline
\end{tabular}

\begin{tablenotes}
\fontsize{10}{15}
\centering
  \item[*] $B(p,q) = \frac{\Gamma(p)\Gamma(q)}{\Gamma(p+q)}$: beta function; $\Gamma(\alpha)$: gamma function.
  \end{tablenotes}
   \end{table}

It should be pointed out that all of these distributions are steady-state distributions of stochastic processes used to describe stochastic volatility. In particular, Ga, IGa and BP are the steady-state distributions of the mean-reverting Heston \cite{heston1993closed,dragulescu2002probability}, multiplicative \cite{nelson1990arch,bouchaud2000wealth}, and combined Heston-multiplicative \cite{dashti2018combined} models respectively. GIGa \cite{ma2013distribution,ma2014model}, GGa and GB2 \cite{hertzler2003classical} are the steady states of non-mean-reverting stochastic processes. Namely, consider a stochastic differential equation.
\begin{equation}
\mathrm{d}x = -\eta(x - \theta x^{1-\alpha})\mathrm{d}t + \sqrt{\kappa_2^2 x^2 + \kappa_\alpha^2 x^ {2-\alpha}}\mathrm{d}W_t
\label{GB2SDE}
\end{equation}
where $\mathrm{d}W_t$ is a Wiener process. Its steady-state distribution is \cite{hertzler2003classical} $GB2(x; p, q,\alpha, \beta)$ in Table \ref{analyticforms} with
\begin{equation}
\label{betaGB2}
\beta=(\frac{\kappa_\alpha}{\kappa_2})^{2 / \alpha}
\end{equation}
\begin{equation}
\label{pGB2}
p=\frac{1}{\alpha}(-1+\alpha +\frac{2 \eta \theta}{\kappa_\alpha^2})
\end{equation}
and
\begin{equation}
\label{qGB2}
q=\frac{1}{\alpha}(1+\frac{2 \eta}{\kappa_2^2})
\end{equation}
The steady-state distribution of (\ref{GB2SDE}) is GIGa for $\kappa_{\alpha}=0$ and GGa for $\kappa_2=0$. For $\alpha=1$ we have mean-reverting models which yield a BP steady-state distribution in general and IGa and Ga for $\kappa_1=0$ and $\kappa_2=0$ respectively.

\subsection{Monthly Data}

\begin{figure}[!htbp]
\centering
\begin{tabular}{c}
\includegraphics[width = 0.49 \textwidth]{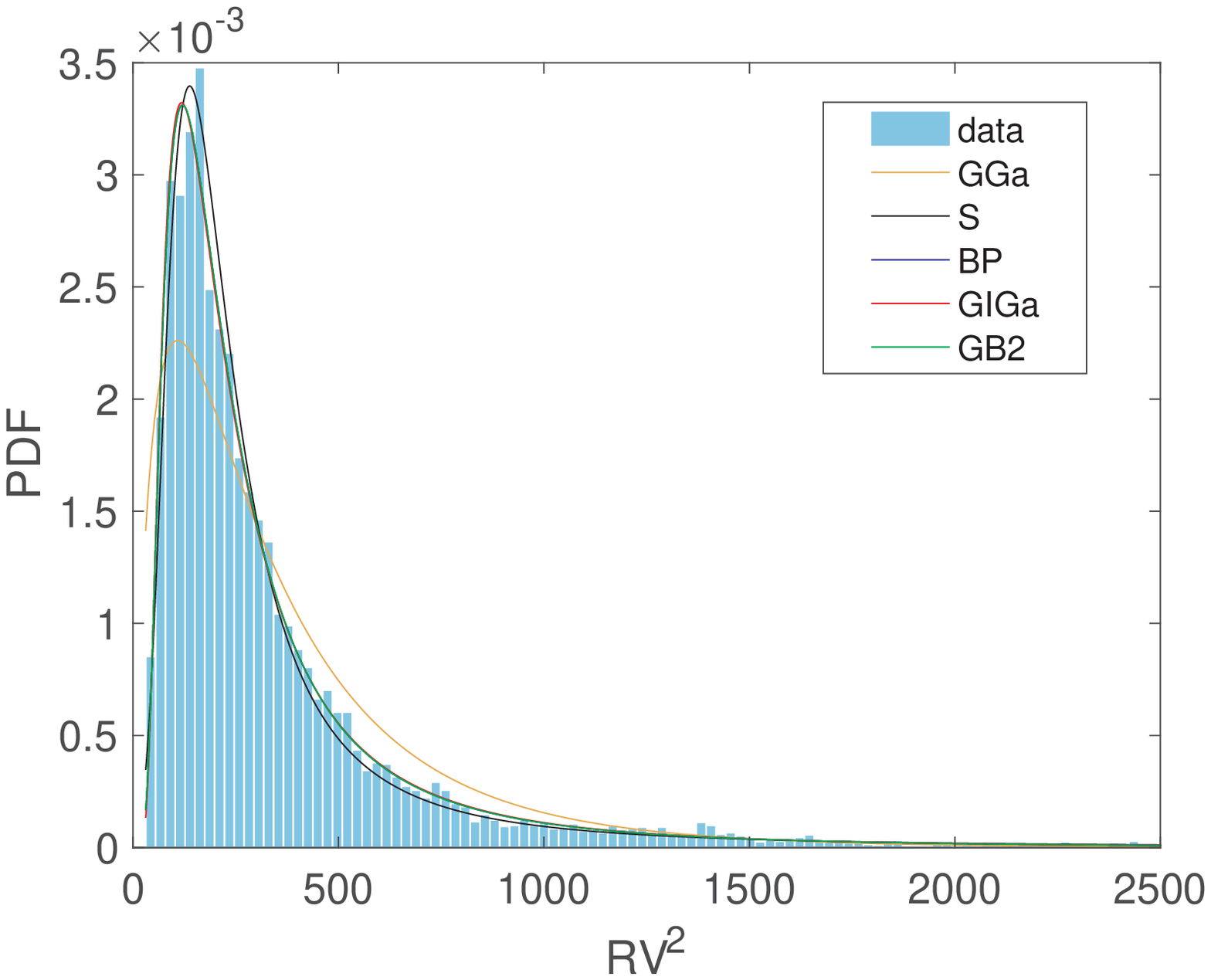}
\includegraphics[width = 0.49 \textwidth]{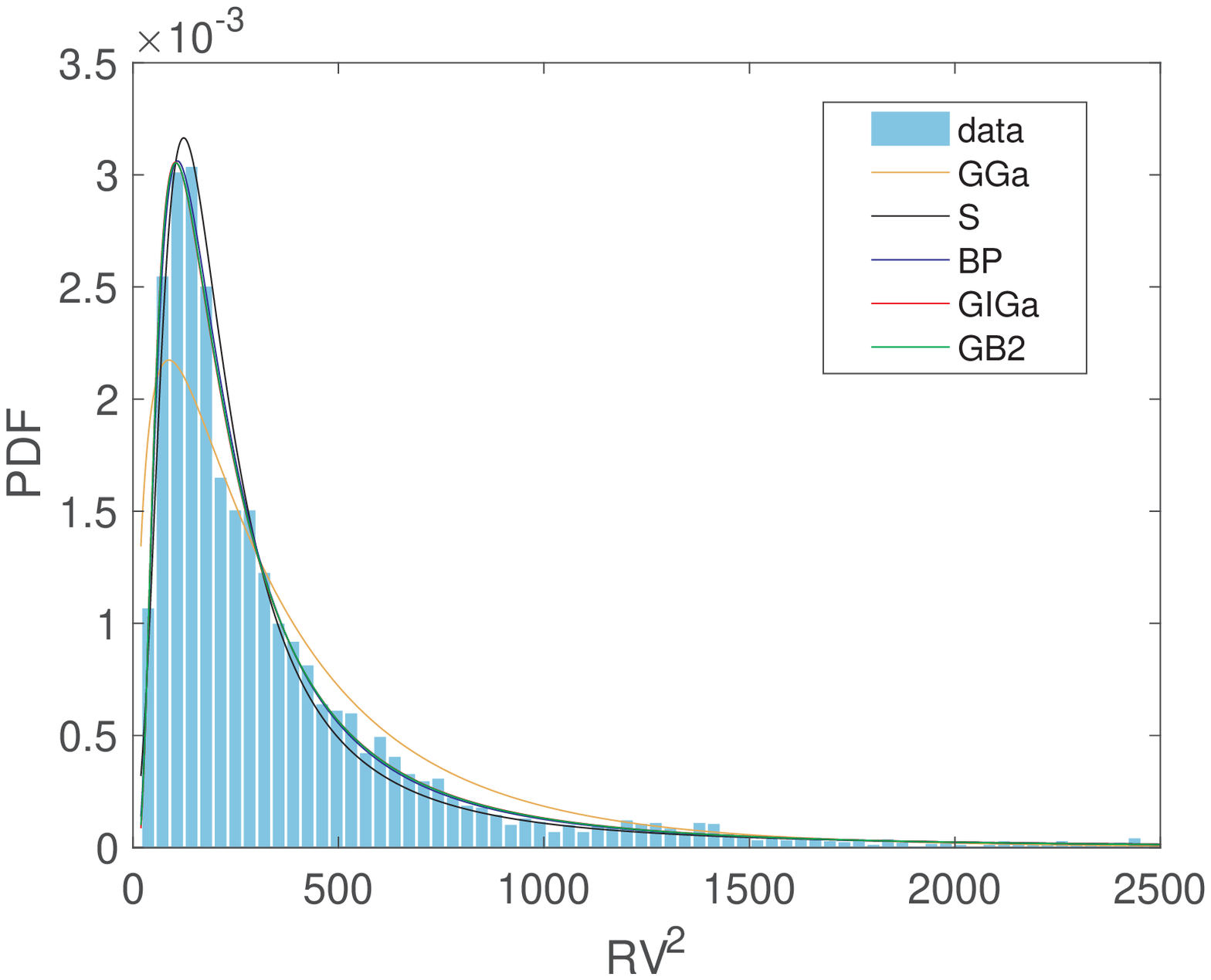}\\
\includegraphics[width = 0.49 \textwidth]{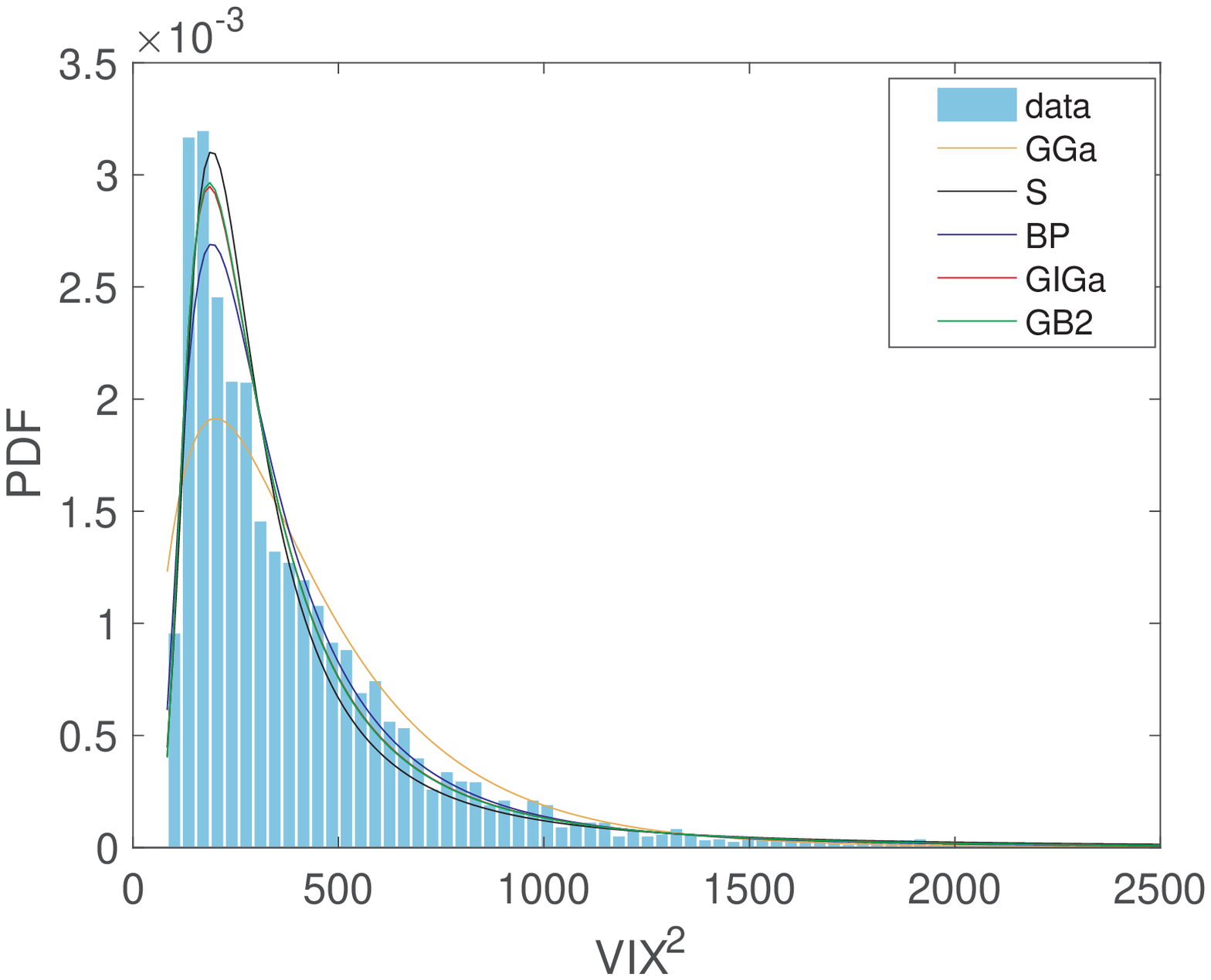}
\includegraphics[width = 0.49 \textwidth]{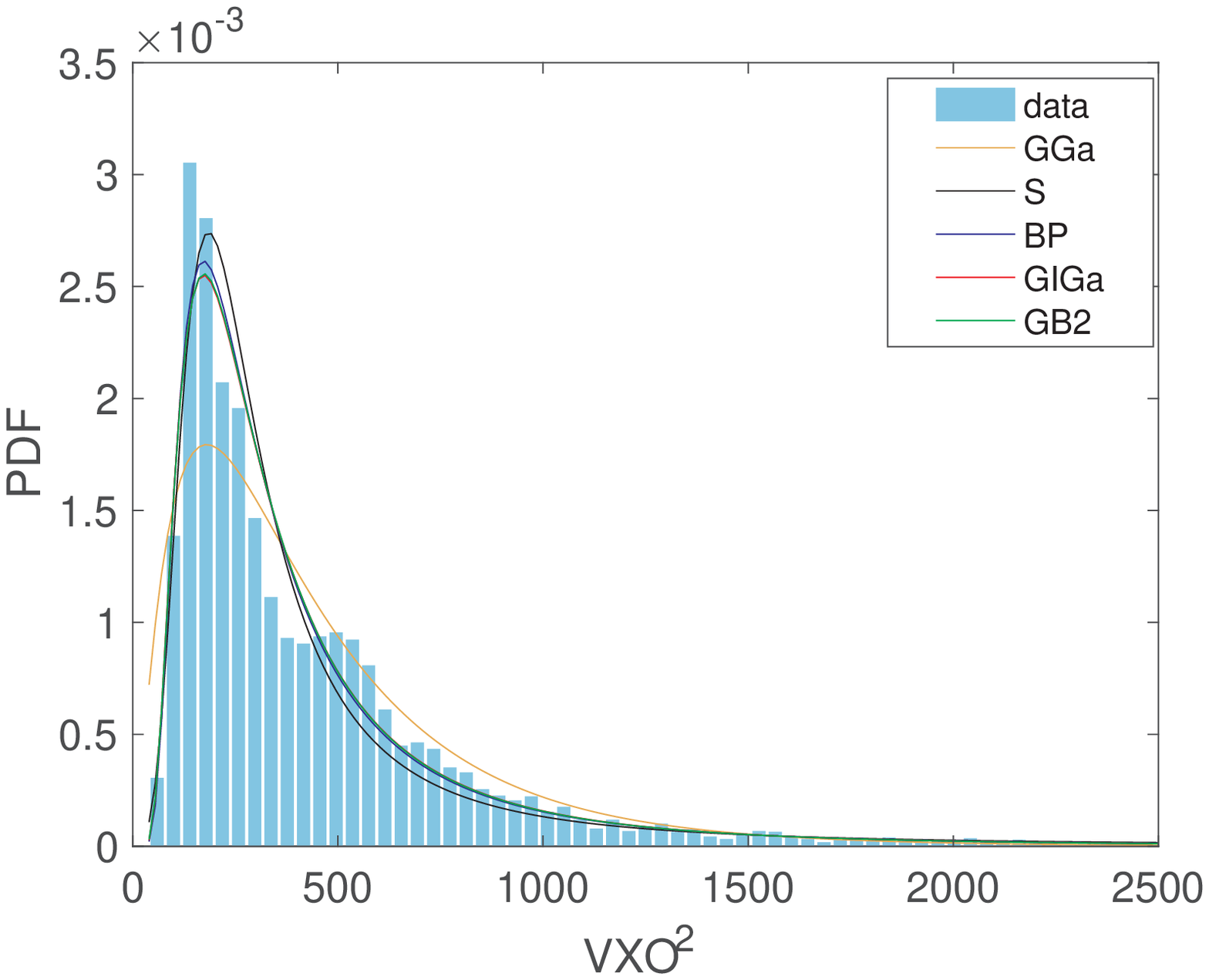}
\end{tabular}
\caption{Clockwise: PDF of monthly $RV^2$ from Jan 2nd, 1970 to Dec 29th, 2017 and PDFs of monthly $RV^2$, $VIX^2$ and $VXO^2$ from Jan 31st, 1990 to Dec 29th, 2017.}
\label{squared}
\end{figure}

Fits of monthly data with the distributions from Table \ref{analyticforms} are shown in Fig. \ref{squared}. Parameters of the distribution fits in Fig. \ref{squared} and their KS statistics are shown in Tables \ref{MLERV22}-\ref{MLEVXO2}. Smaller KS numbers correspond to better fits. For $RV^2$, GB2, BP and GIGa fits are at or close to a 95\% confidence level \cite{knuth1998art}. Obviously, GGa and Ga fit much worse  than any of the fat-tailed distributions. Notice also that the power-law exponents of the front end of GB2 and BP are very large, indicating that the front end is highly suppressed. This explains why GIGa provide nearly as good a fit as GB2. Interestingly, the fits of $VIX^2$ and $VXO^2$ are not nearly as precise as  $RV^2$, which confirms that VIX and VXO are not a very good gauge for predicting RV.
   
\clearpage

\begin{table}[!htbp]
\centering
\caption{MLE results for $RV^2$ from Jan 2nd, 1970 to Dec 29th, 2017}
\label{MLERV22}
\begin{tabular}{ccccc} 
\hline
            type &       parameters&		front exp&		tail exp&          KS test  \\
\hline
Stable & S(          0.9686,           1.0000,           84.0679,           175.8546)  &		 &		 -1.9686&           0.0202 \\
\hline
GB2 & GB2(         15.9183,           1.8735,           1.0150,          23.7045)  &		15.1570&		-2.9016&           0.0115 \\
\hline
BP & BP(         17.2160,           1.9116,           21.9595)  &		16.2160&		-2.9116&           0.0116 \\
\hline
GIGa & GIGa(          2.5562,          625.4491,           0.8023) &		  &		-3.0508 &           0.0138 \\
\hline
IGa & IGa(          1.7394,           319.7392) &		  &		-2.7394&           0.0203 \\
\hline
GGa & GGa(          5.0882,           11.1902 ,    0.4812)&		1.4484  &		 &           0.0786 \\
\hline
Ga & Ga(          1.1391,           364.0363) &		0.1391   &		 &           0.1330 \\
\hline
\end{tabular}
\end{table}

\begin{table}[!htbp]
\centering
\caption{MLE results for $RV^2$ from Jan 31st, 1990 to Dec 29th, 2017}
\label{MLERV222}
\begin{tabular}{ccccc} 
\hline
            type &       parameters&		front exp&		tail exp&          KS test  \\
\hline
Stable & S(          0.9033,           1.0000,          91.7350,         168.7239)  &		 &		-1.9033&           0.0289 \\
\hline
GB2 & GB2(         14.1895,           3.1115,           0.6613,          15.6843) &		8.3835&		-3.0576&           0.0134 \\
\hline
BP & BP(         16.2164,           1.8349,          16.19619) &		15.2164&		-2.8349&           0.0137 \\
\hline
GIGa & GIGa(          3.8505,        2195.2527,          0.5631 ) &		  &		-3.1682&           0.0140 \\
\hline
IGa & IGa(          1.4149,           245.5728) &		 &		-2.4149&           0.0296 \\
\hline
GGa & GGa(          4.2662,           17.0561,    0.4900)&		1.0904  &		 &           0.0652 \\
\hline
Ga & Ga(          1.0295,           436.8326)&		0.0295  &		 &           0.1163 \\
\hline
\end{tabular}
\end{table}

\begin{table}[!htbp]
\centering
\caption{MLE results for $VIX^2$ from Jan 31st, 1990 to Dec 29th, 2017}
\label{MLEVIX2}
\begin{tabular}{ccccc} 
\hline
            type &       parameters&		front exp&		tail exp&          KS test  \\
\hline
Stable & S(          0.9548,           1.0000,           92.1996,           234.4670) &			&		-1.9548&           0.0486 \\
\hline
GB2 & GB2(         63.3797,           1.3249,           1.4751,		18.1068) &		92.5294&		-2.9544 &           0.0363 \\
\hline
BP & BP(         44.1482,           2.6245,           16.1142) &		43.1482&		-3.6245 &           0.0407 \\
\hline
GIGa & GIGa(          1.4520,           325.9344,           1.3814) &		 &		-3.0058&           0.0375 \\
\hline
IGa & IGa(          2.5156,           667.9832)&		  &		-3.5156&           0.0402 \\
\hline
GGa & GGa(         6.8529,          3.1634,           1.0607) &           6.2689  &		 &           0.0693 \\
\hline
Ga & Ga(          1.8988,           230.0093) &           0.8988 &		  &           0.0882 \\
\hline
\end{tabular}
\end{table}

\begin{table}[!htbp]
\centering
\caption{MLE results for $VXO^2$ from Jan 31st, 1990 to Dec 29th, 2017}
\label{MLEVXO2}
\begin{tabular}{ccccc} 
\hline
            type &       parameters&		front exp&		tail exp&          KS test  \\
\hline
Stable & S(          0.9554,           1.0000,           104.3782,           232.7193) &		 &		-1.9554&           0.0564 \\
\hline
GB2 & GB2(         58.4930,           2.6432,           0.8839,		8.1216)  &		50.7020&		-3.3363&           0.0392 \\
\hline
BP & BP(         44.1507,           2.1309,           12.5195) &		43.1507&		-3.1309&           0.0401 \\
\hline
GIGa & GIGa(          3.0721,          1092.4445,           0.7954)&		  &		-3.4435 &           0.0423 \\
\hline
IGa & IGa(          2.0448,           519.0907) &		 &		-3.0048 &           0.0483 \\
\hline
GGa & GGa(         5.599,           16.0437,           0.5327) &		1.9826 &		  &           0.0713 \\
\hline
Ga & Ga(          1.6328,           283.4215) &		0.6328 &		  &           0.0922 \\
\hline
\end{tabular}
\end{table}

Aside from an obvious qualitative difference between $RV^2$ 1970-2017 and $RV^2$ 1990-2017 in Fig. \ref{squared}, we observe that for GB2 fitting the front end (low volatilities) of the former is significantly more suppressed than for the latter, which, incidentally, is also true for $VIX^2$ relative to $VXO^2$. The tail exponents, on the other hand, are much closer to each other.

\subsection{Development of $RV^2$ Distribution as Function of Number of Days}

\begin{figure}[!htbp]
\centering
\begin{tabular}{c}
\includegraphics[width = 0.56 \textwidth]{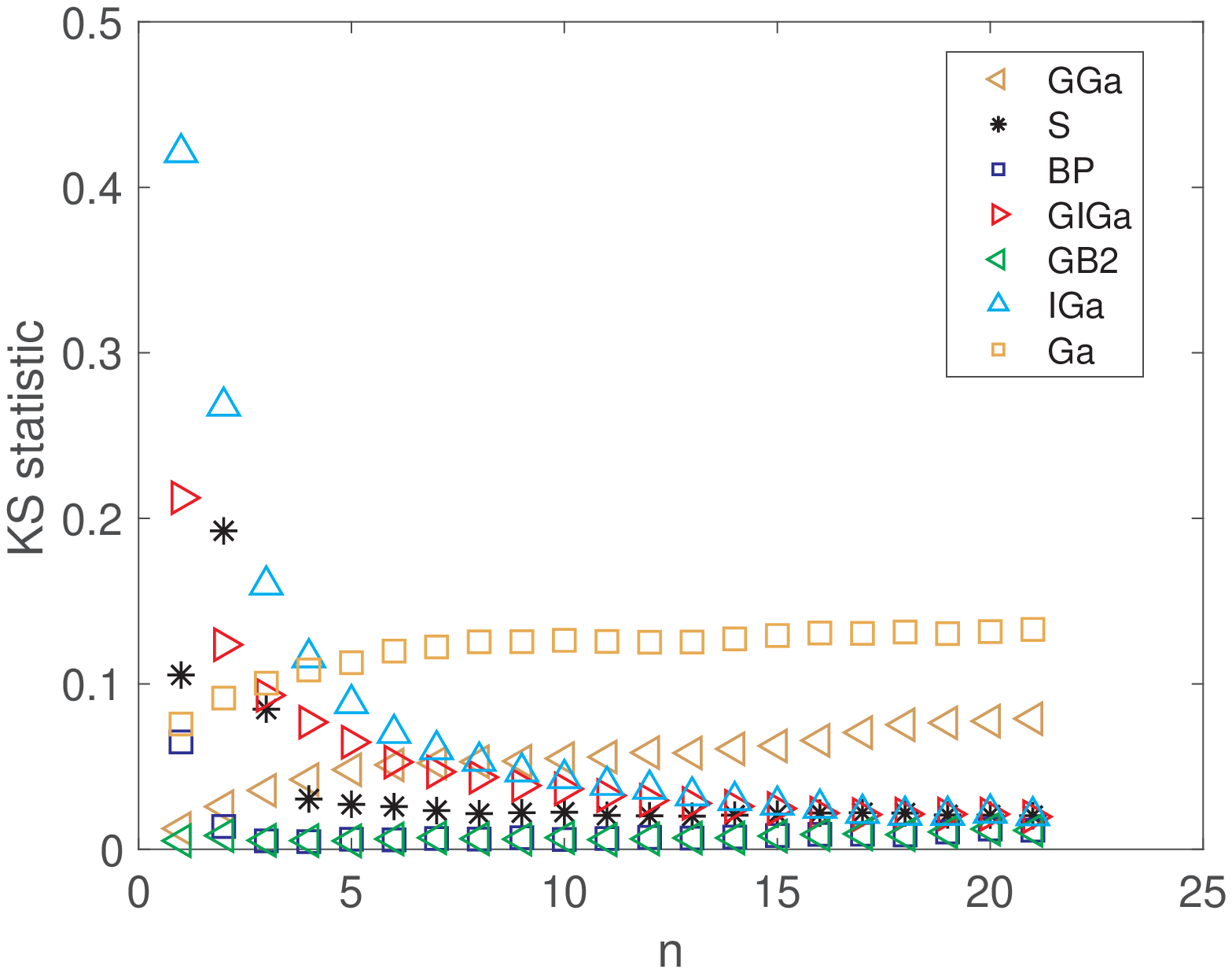}\\
\includegraphics[width = 0.56 \textwidth]{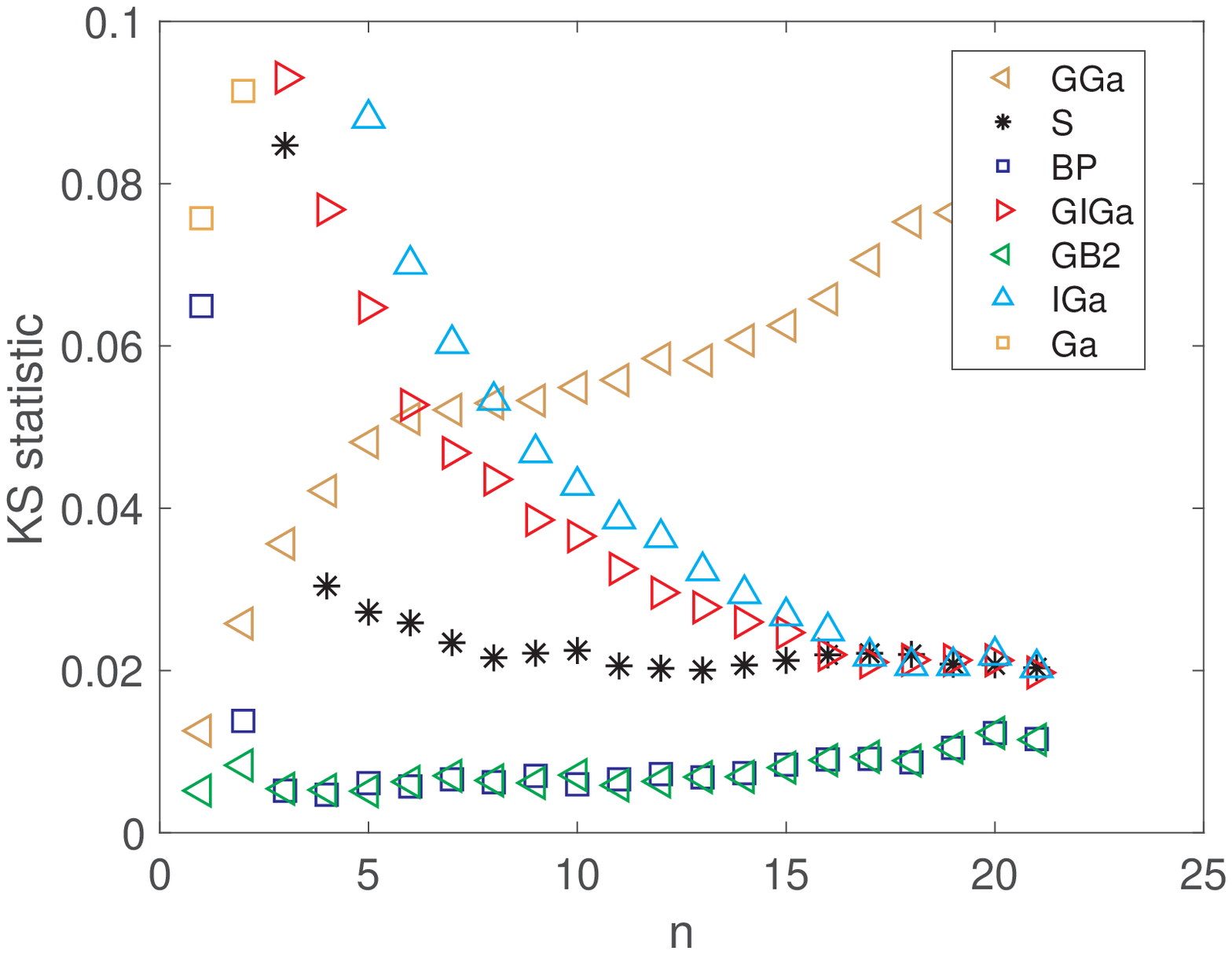}
\end{tabular}
\caption{KS statistics as function of $n$. Top and bottom graphs are the same but for a different vertical scale.}
\label{KS-n}
\end{figure}

\begin{figure}[!htbp]
\centering
\begin{tabular}{c}
\includegraphics[width = 0.35 \textwidth]{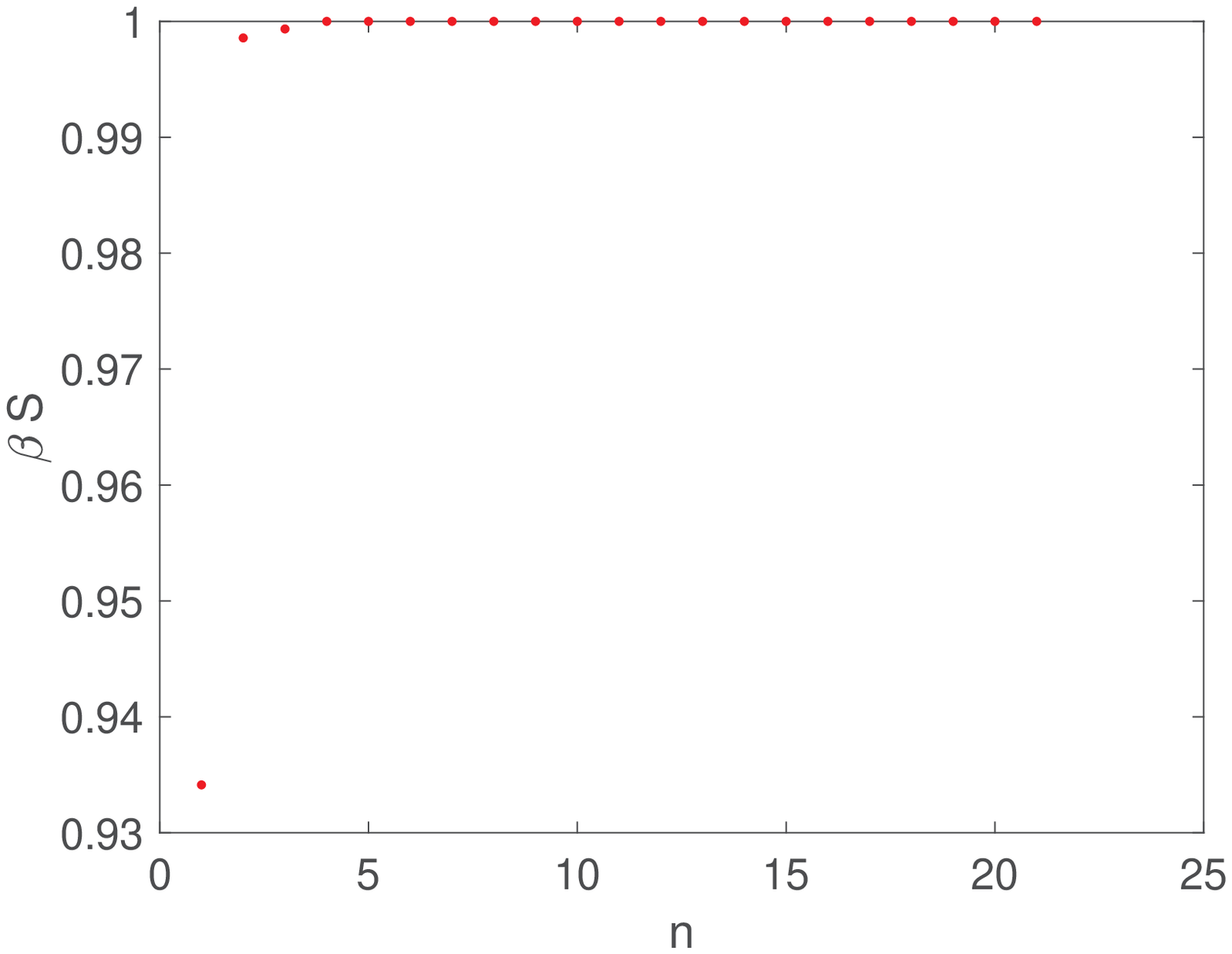}
\includegraphics[width = 0.35 \textwidth]{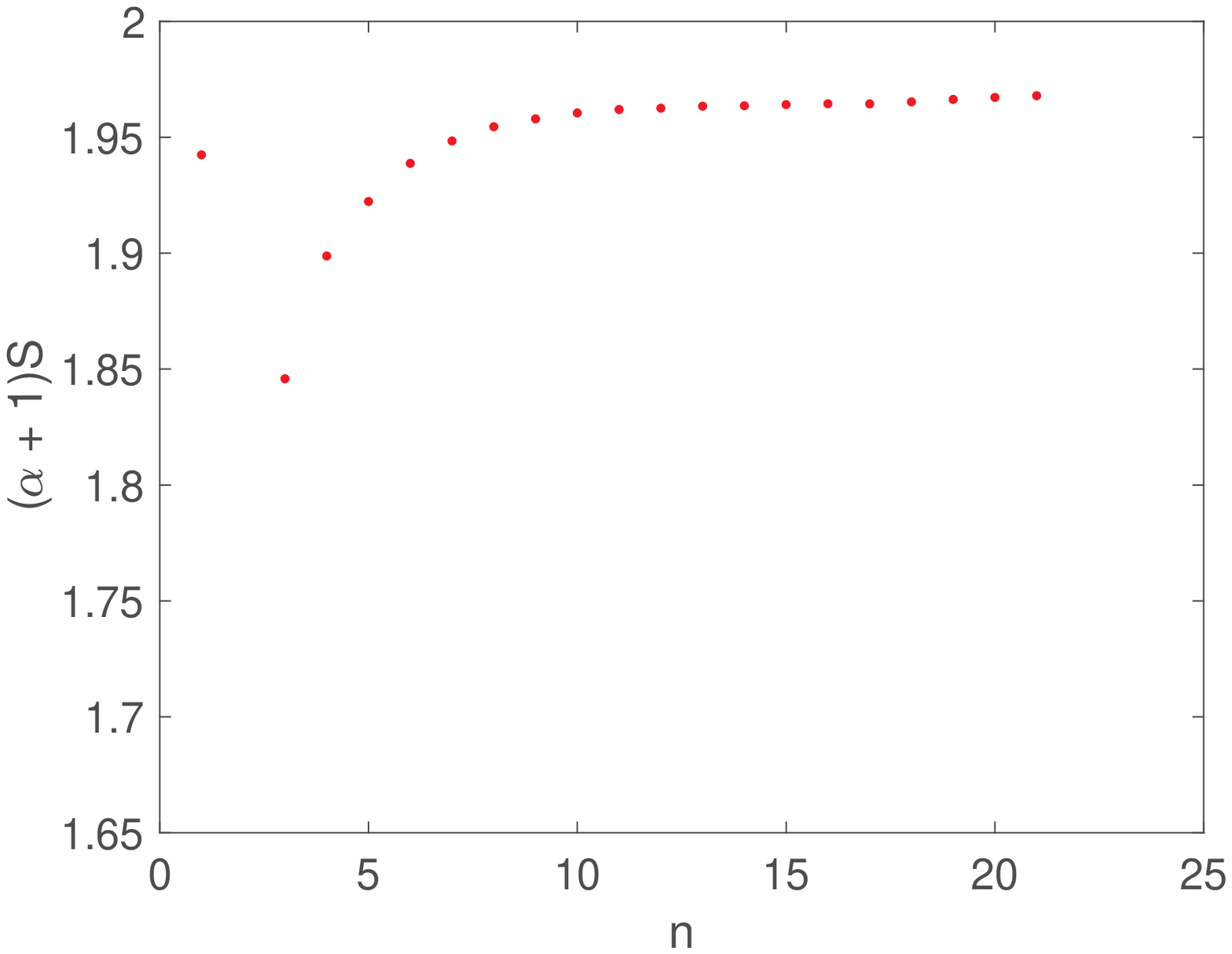}\\
\includegraphics[width = 0.35 \textwidth]{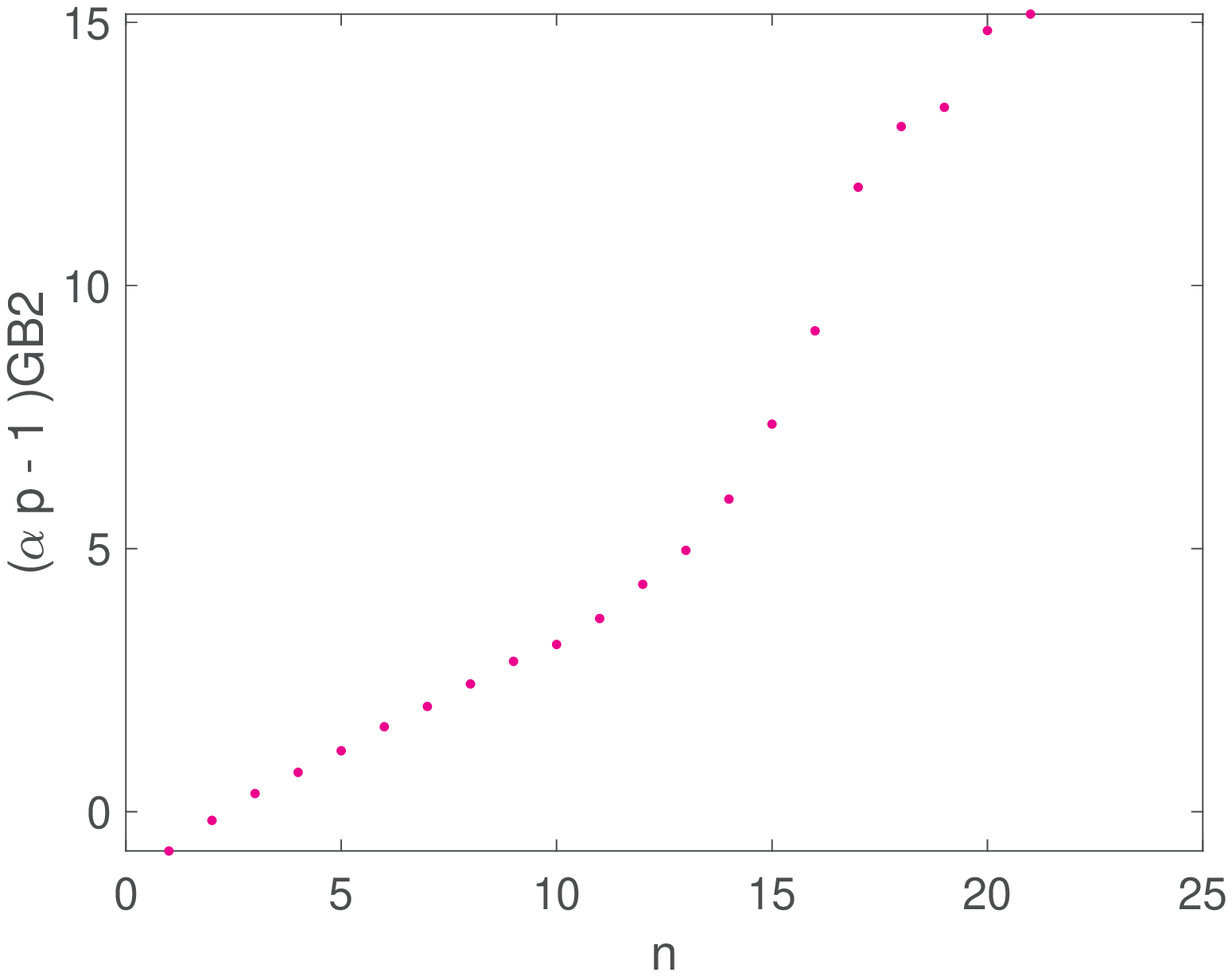}
\includegraphics[width = 0.35 \textwidth]{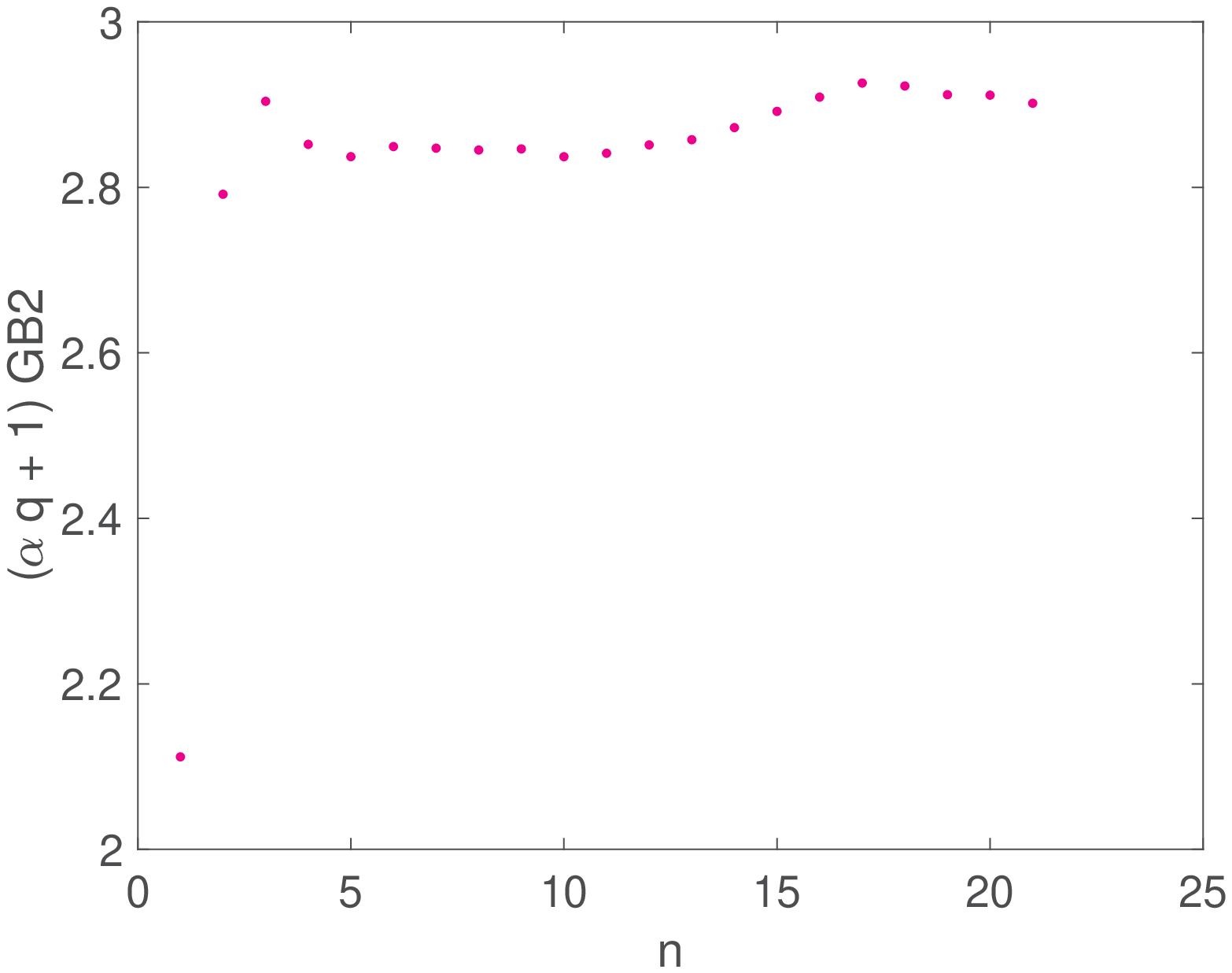}\\
\includegraphics[width = 0.35 \textwidth]{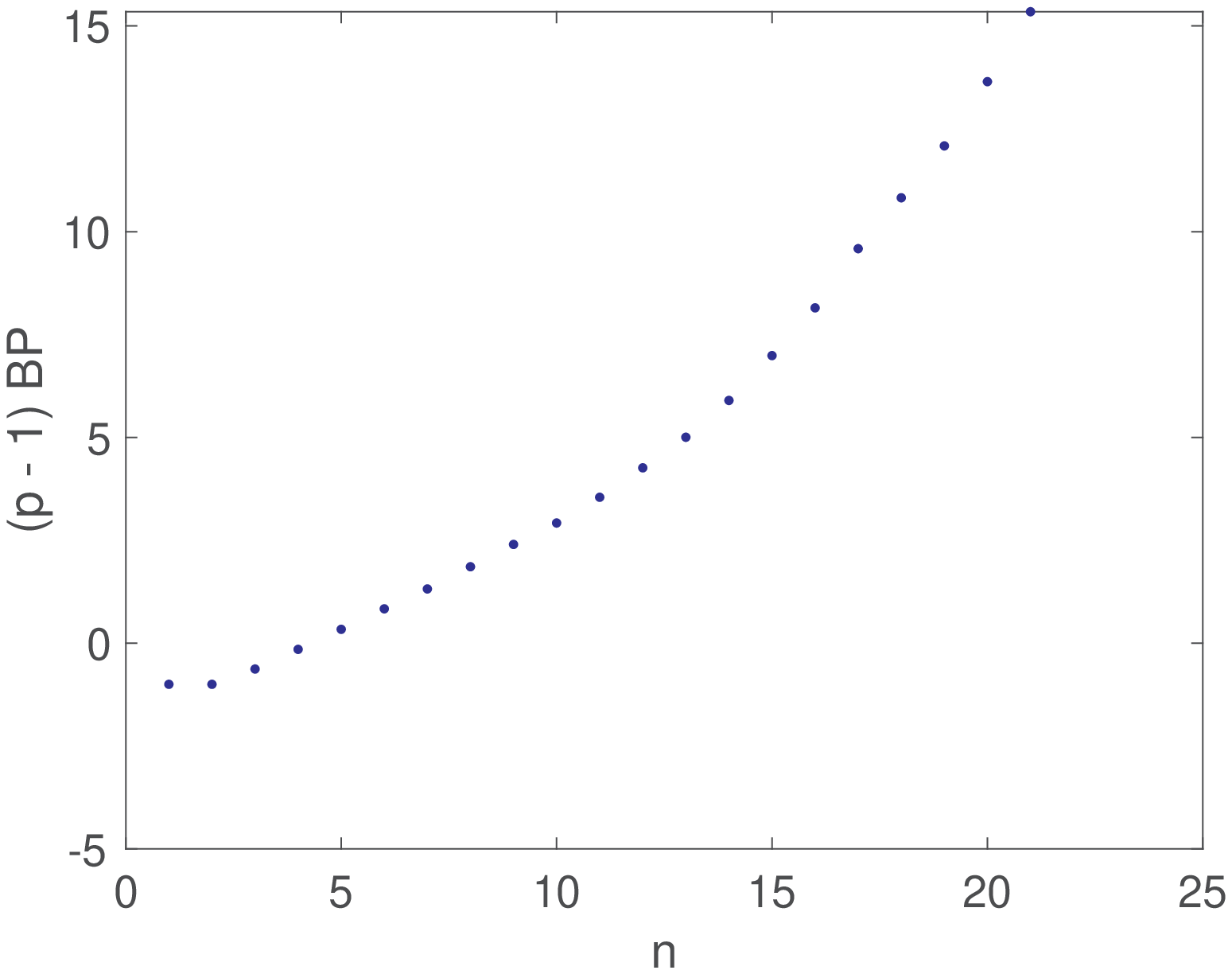}
\includegraphics[width = 0.35 \textwidth]{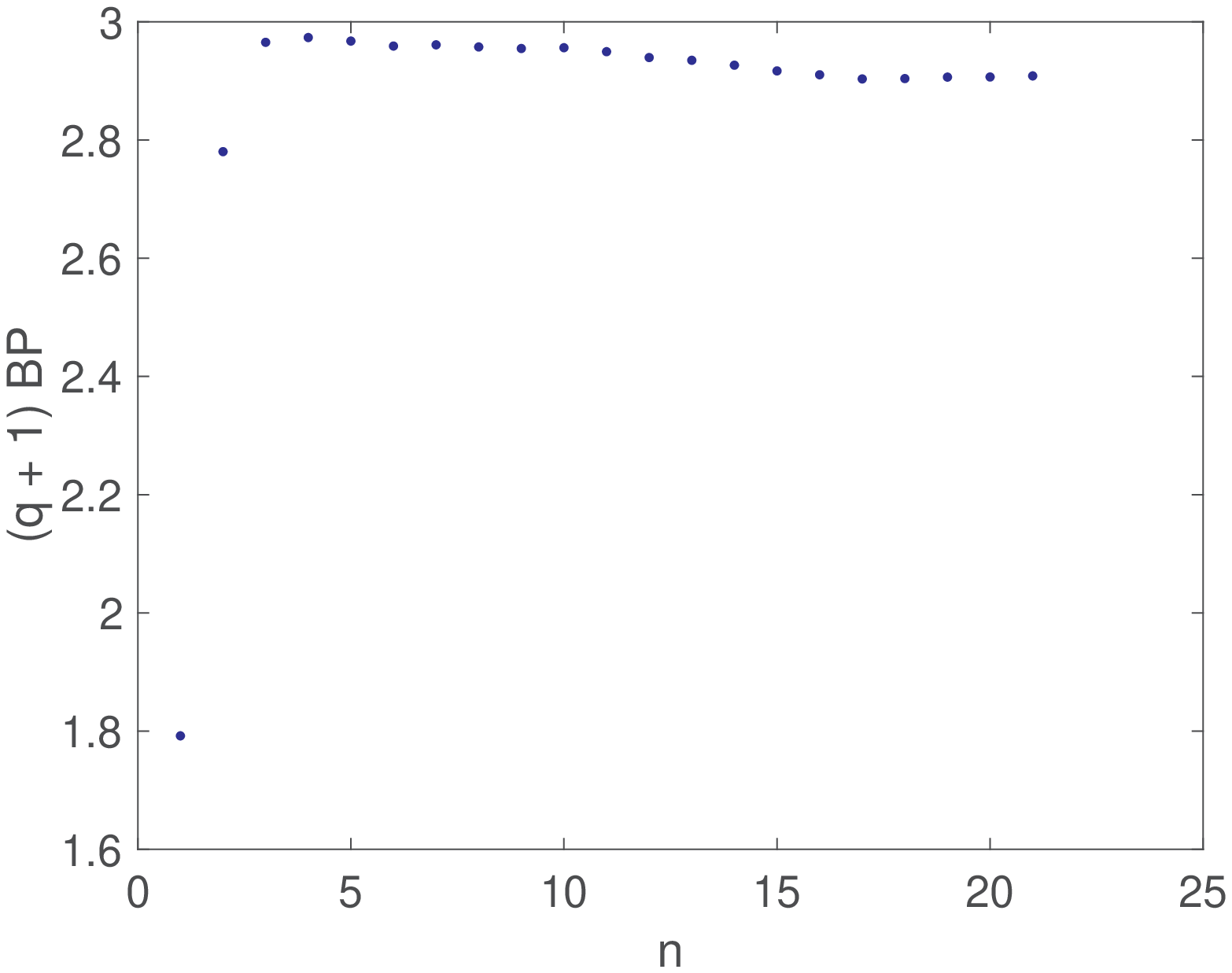}\\
\includegraphics[width = 0.35 \textwidth]{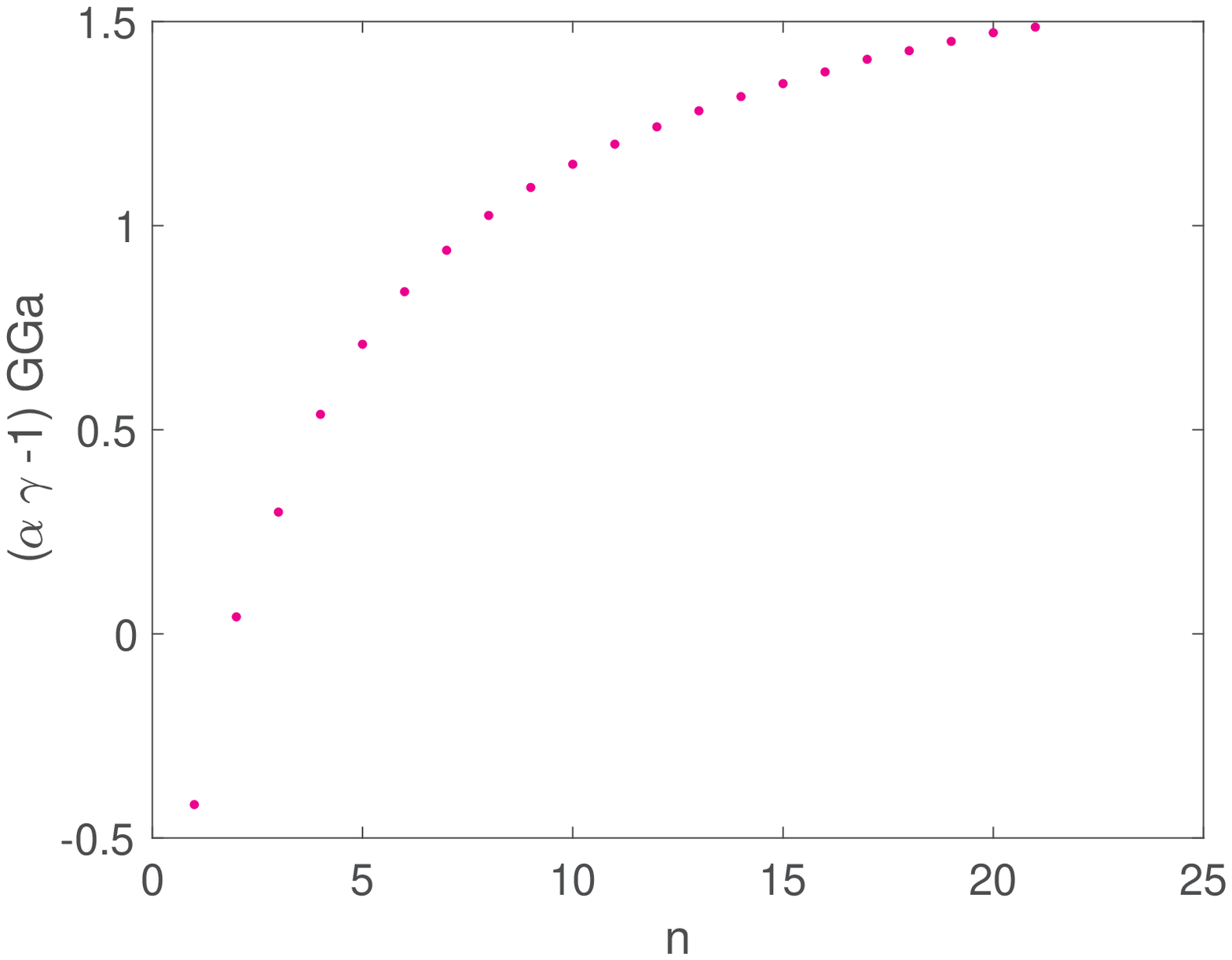}
\includegraphics[width = 0.35 \textwidth]{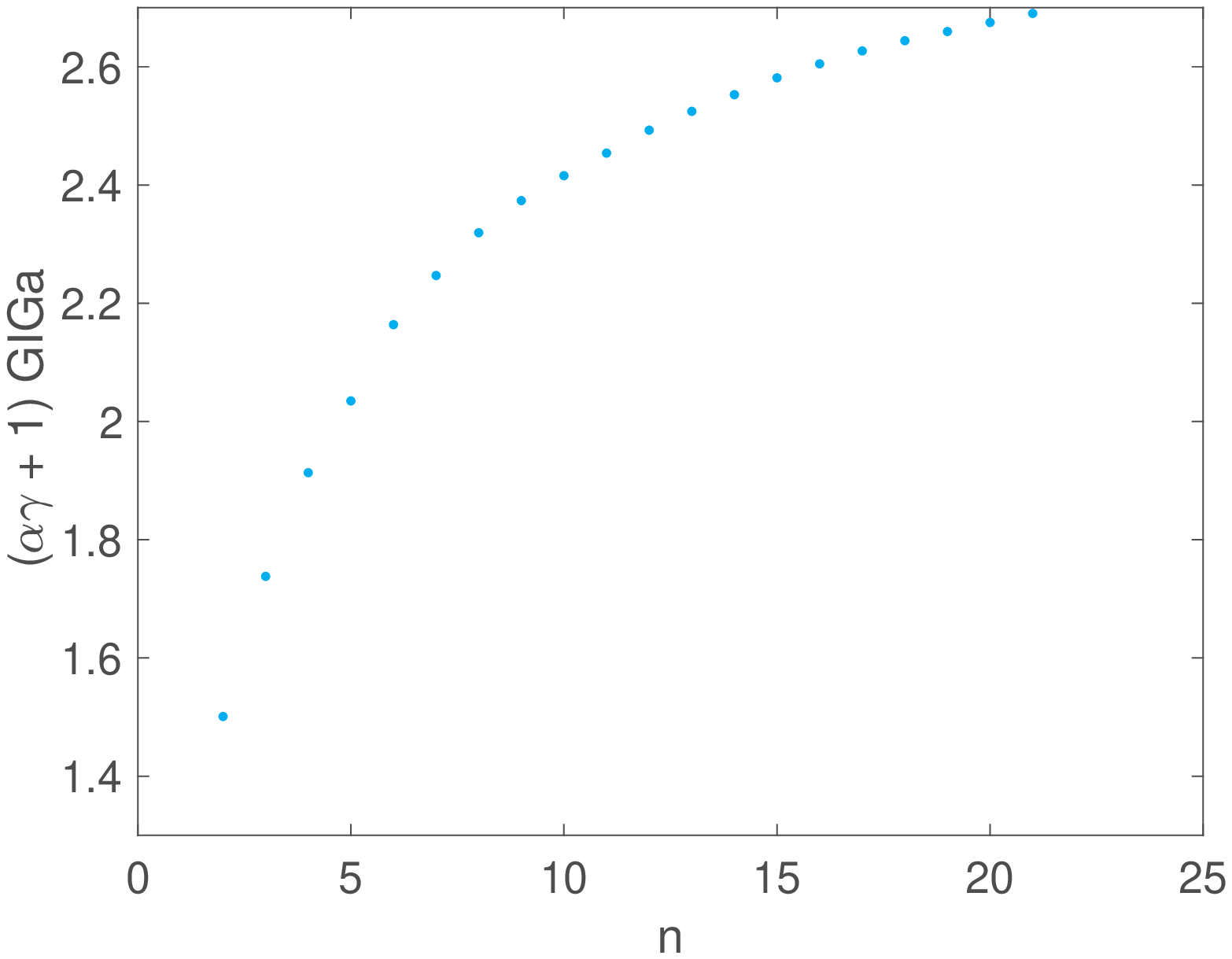}
\end{tabular}
\caption{Power-law exponents, as per Table \ref{analyticforms} as a function of $n$.}
\label{parameters-n}
\end{figure}

\begin{figure}[!htbp]
\centering
\begin{tabular}{c}
\includegraphics[width = 0.35 \textwidth]{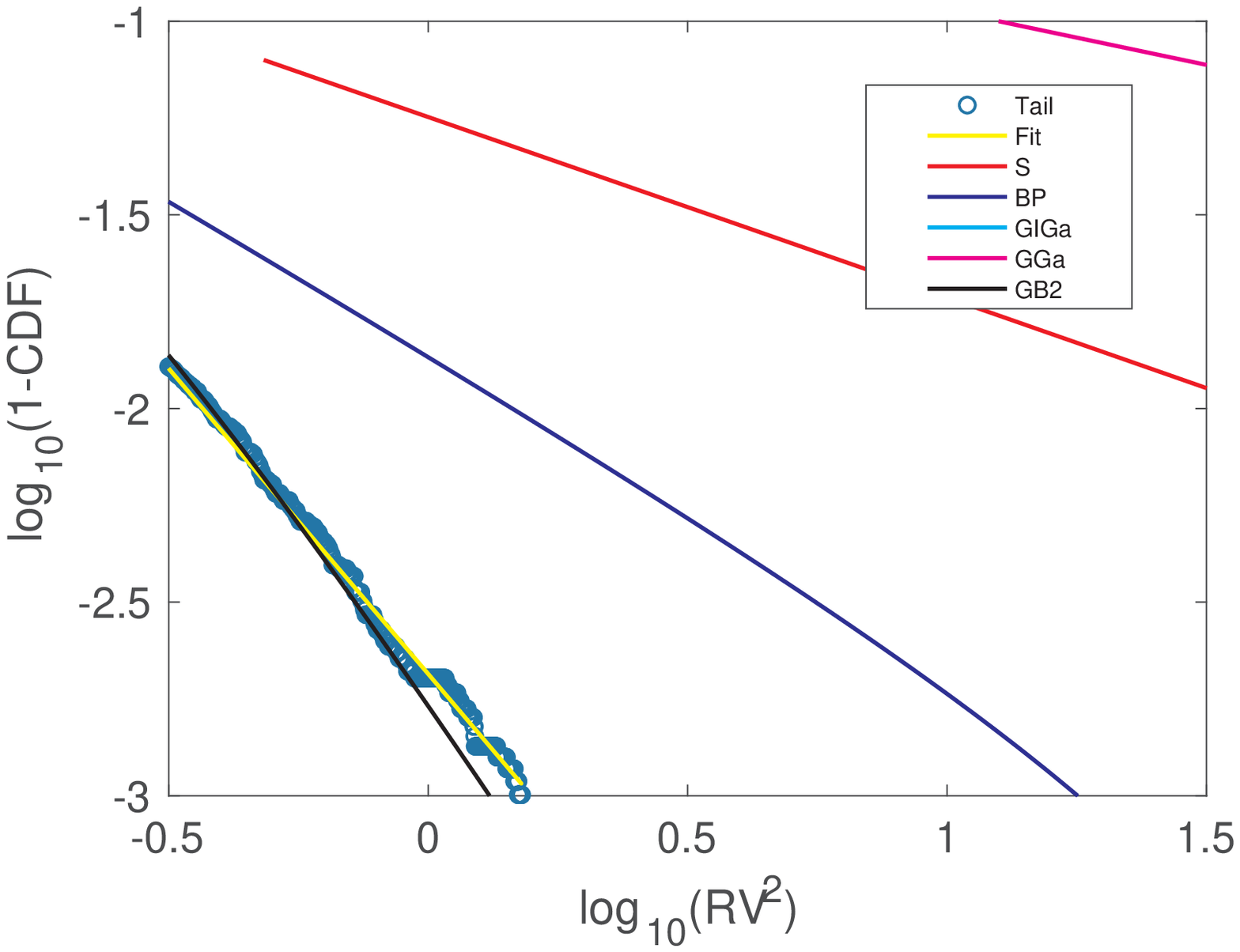}
\includegraphics[width = 0.35 \textwidth]{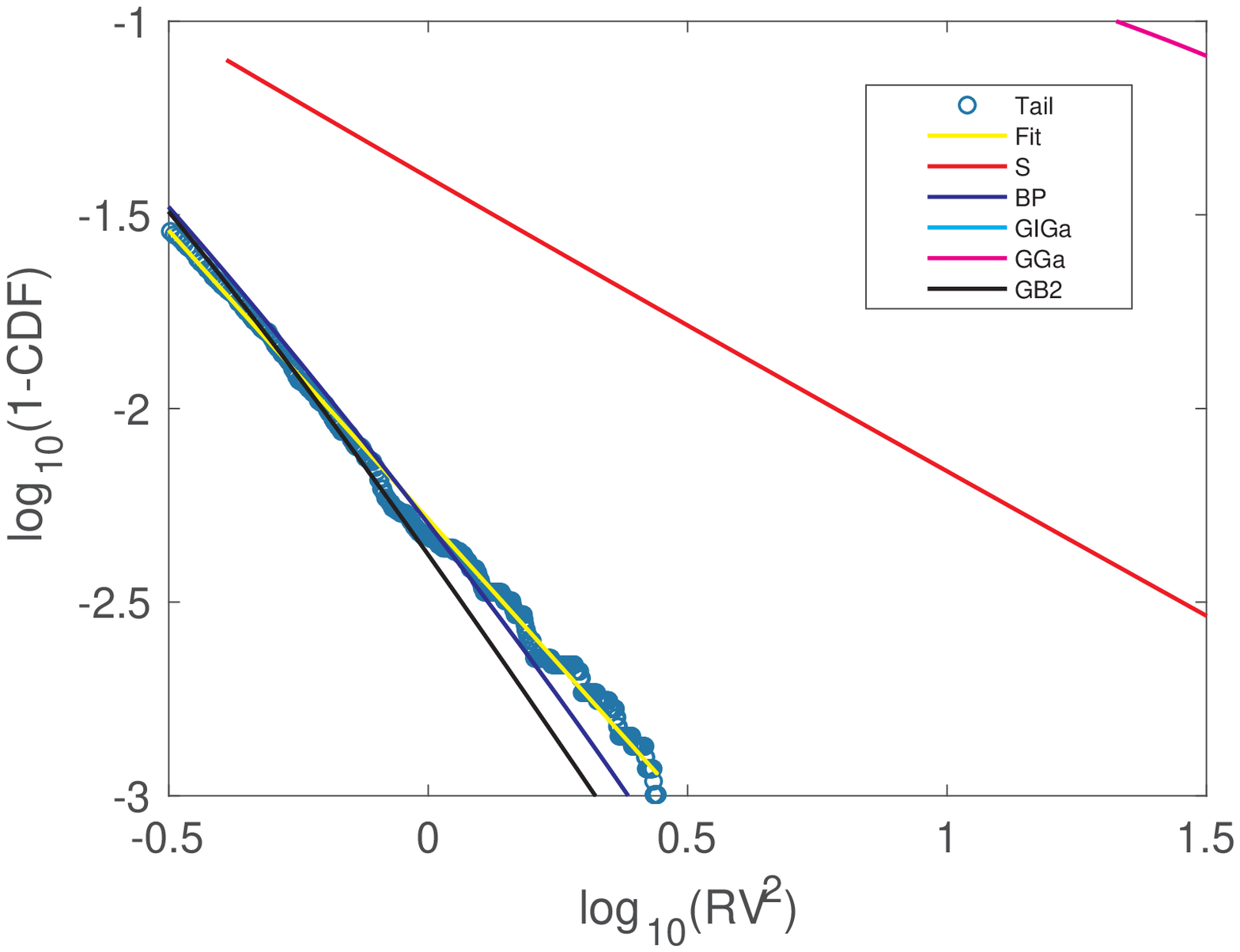}\\
\includegraphics[width = 0.35 \textwidth]{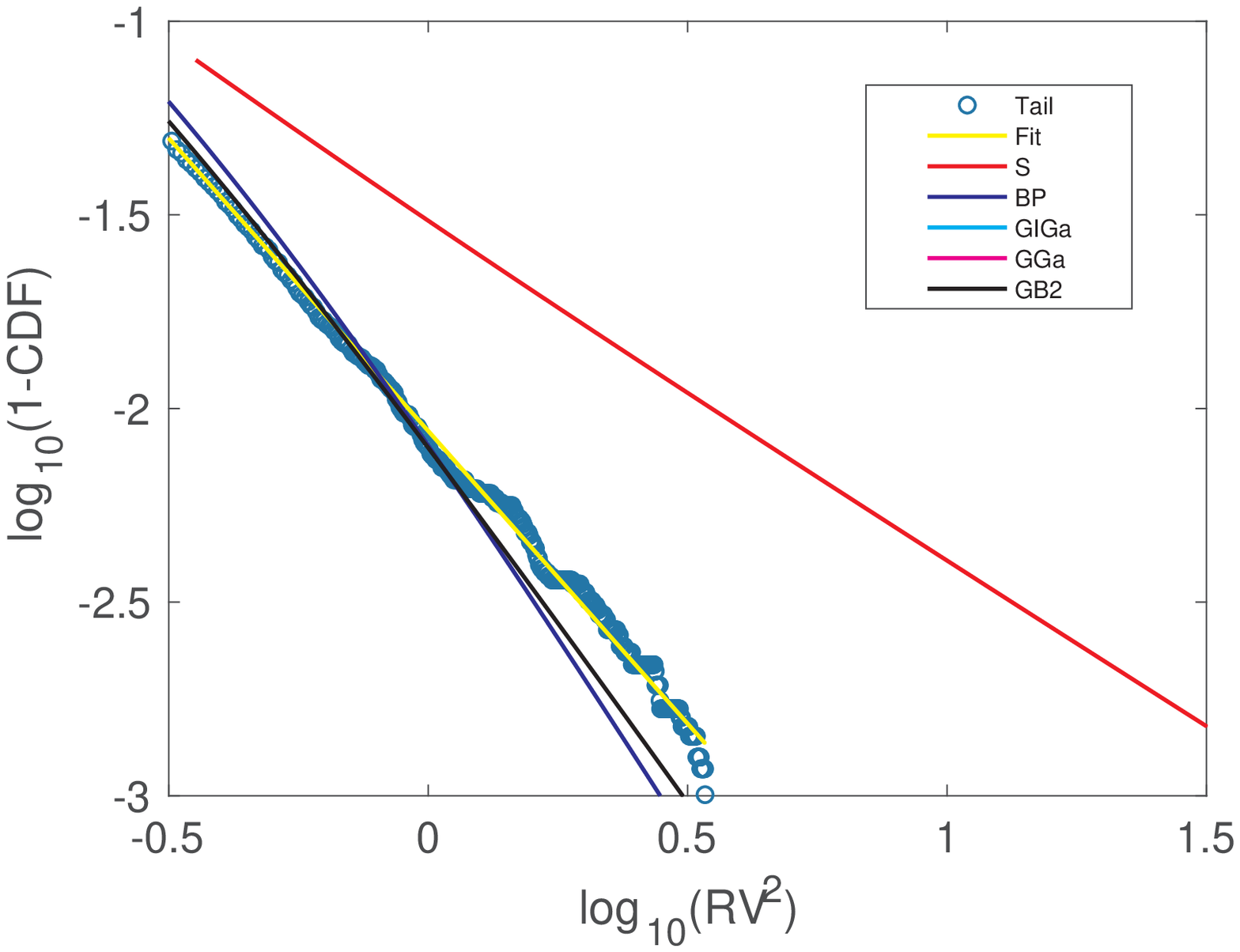}
\includegraphics[width = 0.35 \textwidth]{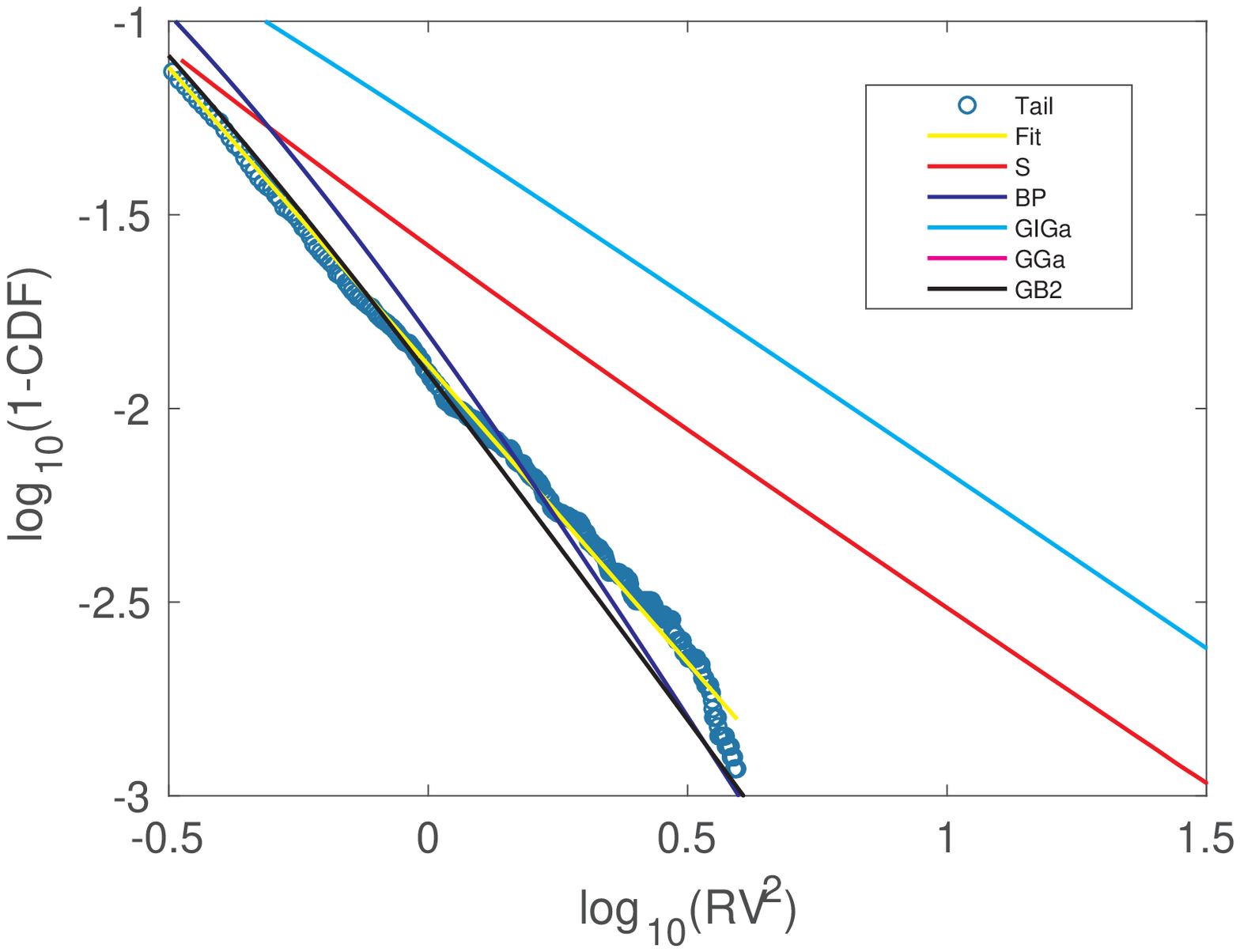}\\
\includegraphics[width = 0.35 \textwidth]{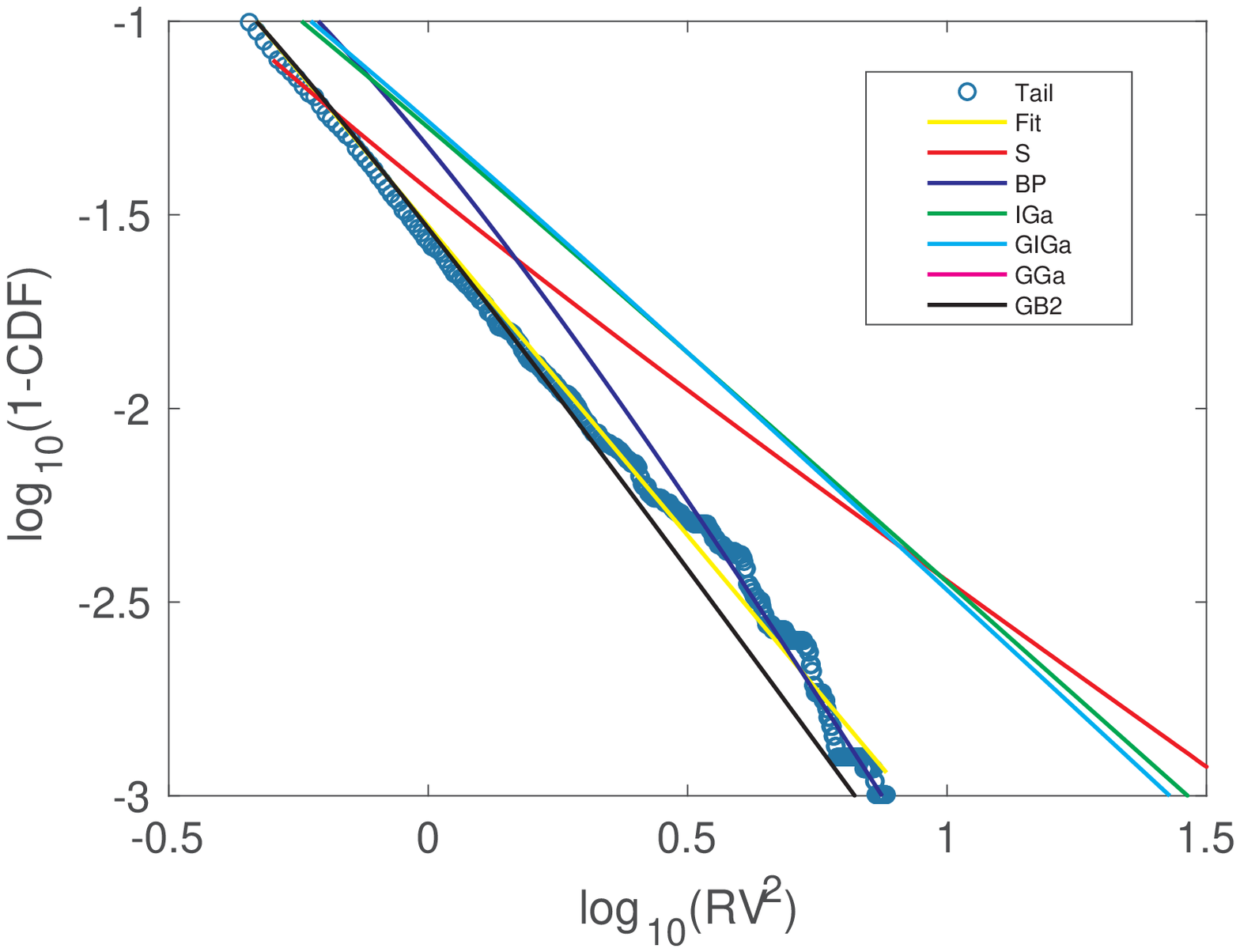}
\includegraphics[width = 0.35 \textwidth]{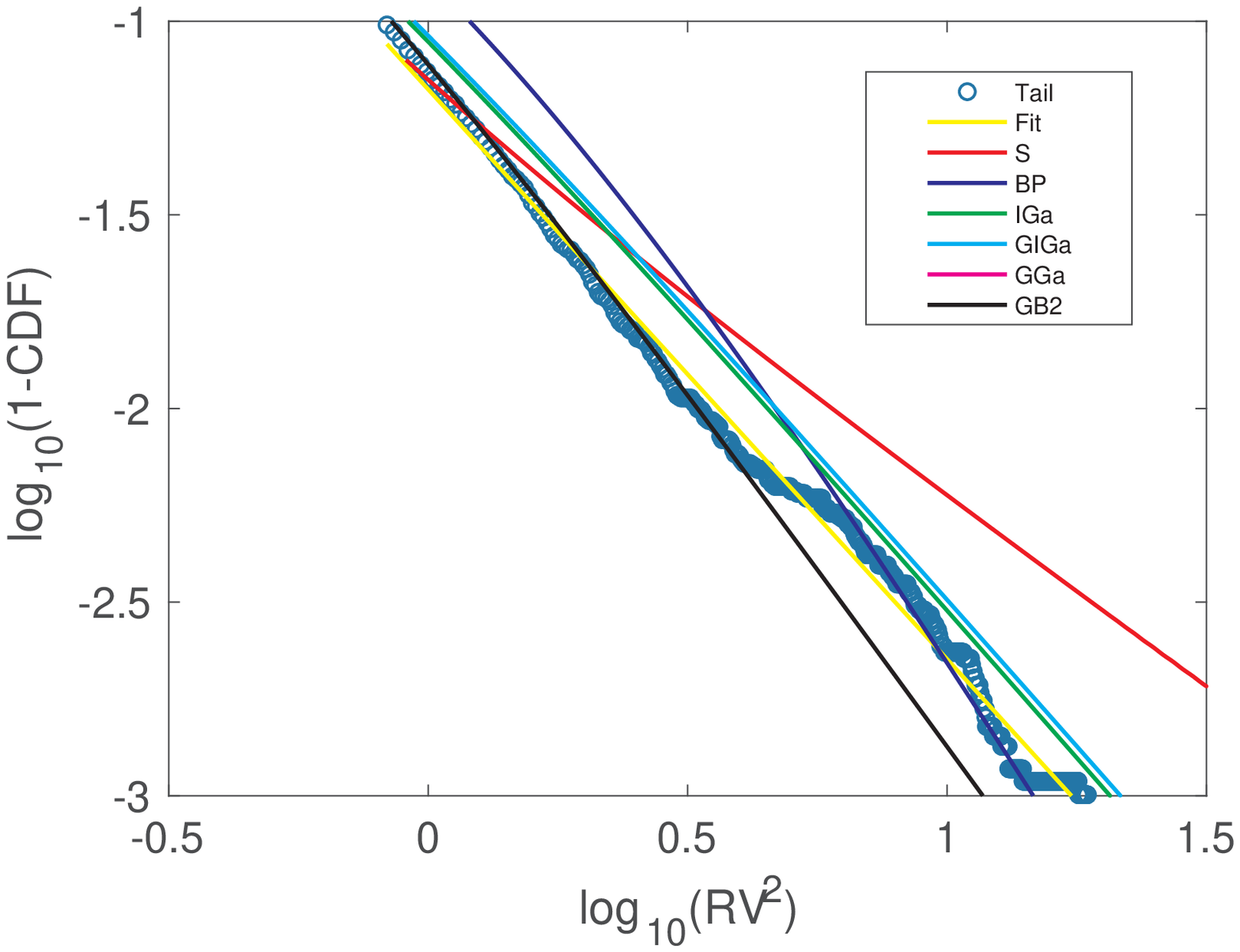}\\
\includegraphics[width = 0.35 \textwidth]{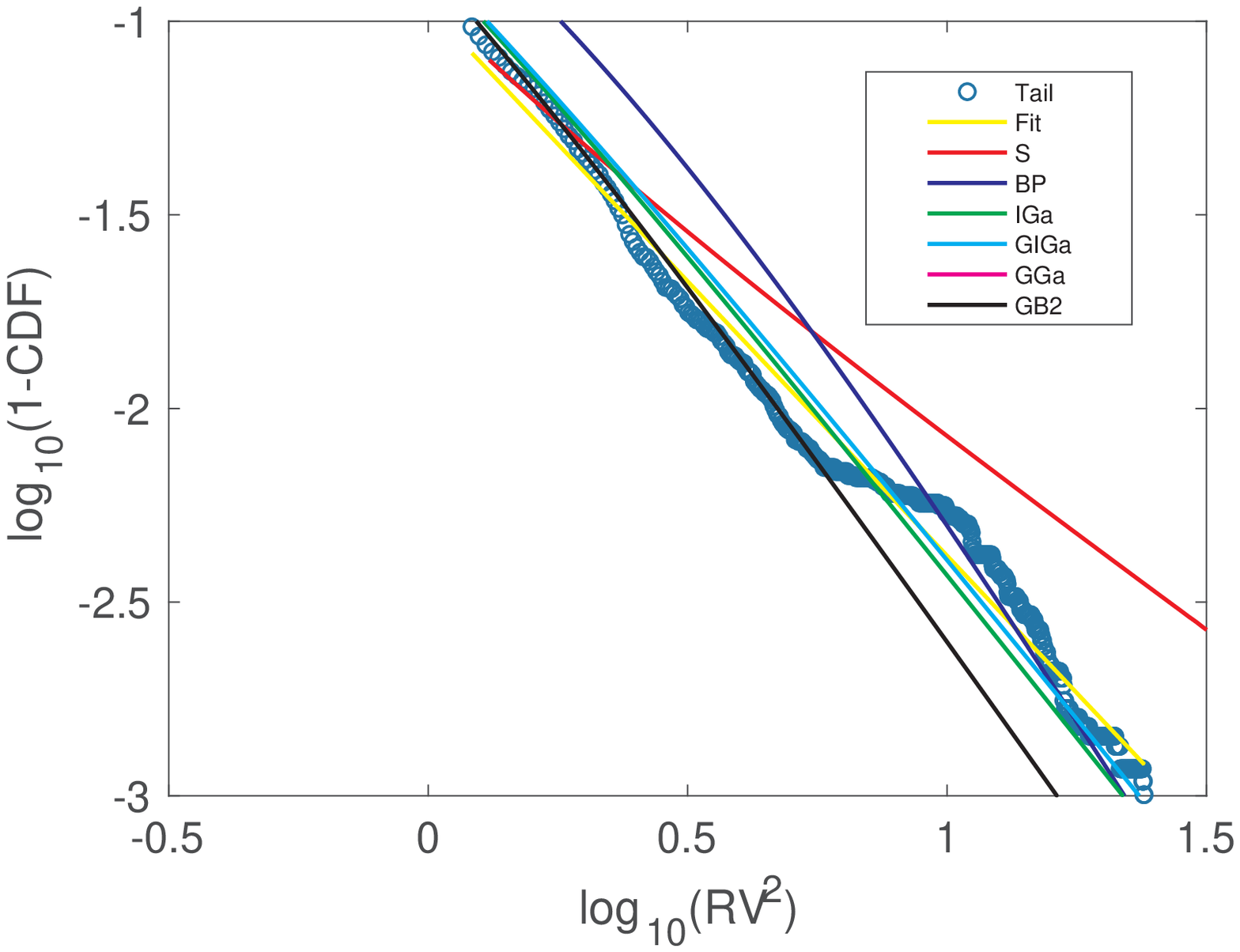}
\includegraphics[width = 0.35 \textwidth]{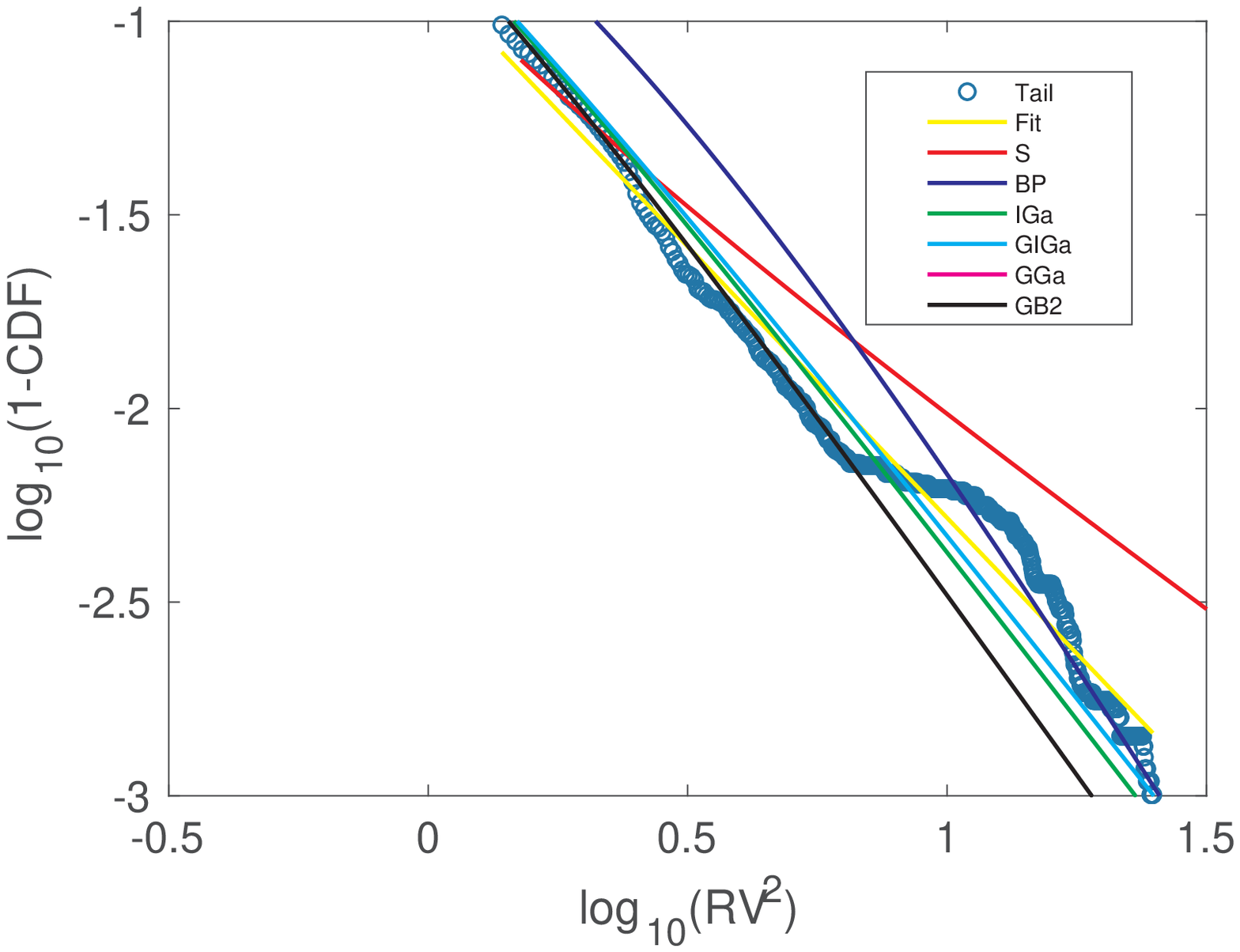}
\end{tabular}
\caption{Tails of fitted distribution vis-a-vis the actual tail and its linear fit as a function of $n$. From left to right and top to bottom, the plots are for $n=$  1, 2, 3, 4, 6, 12, 18, 21 days.}
\label{tails-n}
\end{figure}

Fig. \ref{sumri2} shows that the distribution function of $RV^2$ develops rapidly with the number of days $n$ at about $n \approx 5-7$. Here, we take a more careful look at how the parameters of the distribution fits depend on $n$. Fig. \ref{KS-n} gives the $n$-dependence of KS statistics, which compares the goodness of fits. Fig. \ref{parameters-n}, shows the $n$-dependence of power-law exponents. Fig. \ref{tails-n} compares tails of fitted distributions to the actual tail and its fit. The important observations are as follows:
\begin{itemize}
\item{Gap between GIGa/IGa and GB2/BP KS decreases with $n$, as front exponents of the latter grows.}
\item{Front exponents are negative for GB2/BP and GGa for daily $RV^2$, reflecting the absence of bell shape.}
\item{Unlike GIGa/IGa, for S and GB2/BP tail exponents saturate rapidly from smaller (fatter) daily $RV^2$.}
\end{itemize}

Data represented in Figs. \ref{KS-n}, \ref{parameters-n} and \ref{tails-n} reflects the 1970-2017 period but the 1990-2017 subset looks quite similar. Notice that, unlike other distributions here, GGa does not have a heavy tail. Notice also that not all of the distributions ``make it" into every tail-fitting windows in Fig \ref{tails-n}. 

\clearpage
\section {Probability Distribution Functions of Differences} \label{difference}

There is a simple relationship between correlation $\rho$ between two time series $a_i$ and $b_i$ and their rms and rms of the distribution of the difference:
\begin{equation}
\sigma_{a-b}^2=\sigma_a^2 + \sigma_b^2 - 2 \rho \sigma_a \sigma_b
\label{auto}
\end{equation}
Of course, knowledge of the distribution functions of $RV^2$, $VIX^2$, $VXO^2$ and that of the difference of $VIX^2$ and $VXO^2$ with scaled $RV^2$ give you far richer information than simple extractions of correlation coefficient between the indices. Therefore, in this Section we study the distribution of the differences of $VIX^2$ and $VXO^2$ with $RV^2$, where the latter is rescaled per the ratios in Table \ref{ratios} (In other words, ``$VIX^2 - RV^2$" in actuality means $VIX^2 - (mean(VIX^2)/mean(RV^2))RV^2$). 

\begin{figure}[!htbp]
\centering
\begin{tabular}{cc}
\includegraphics[width = 0.49 \textwidth]{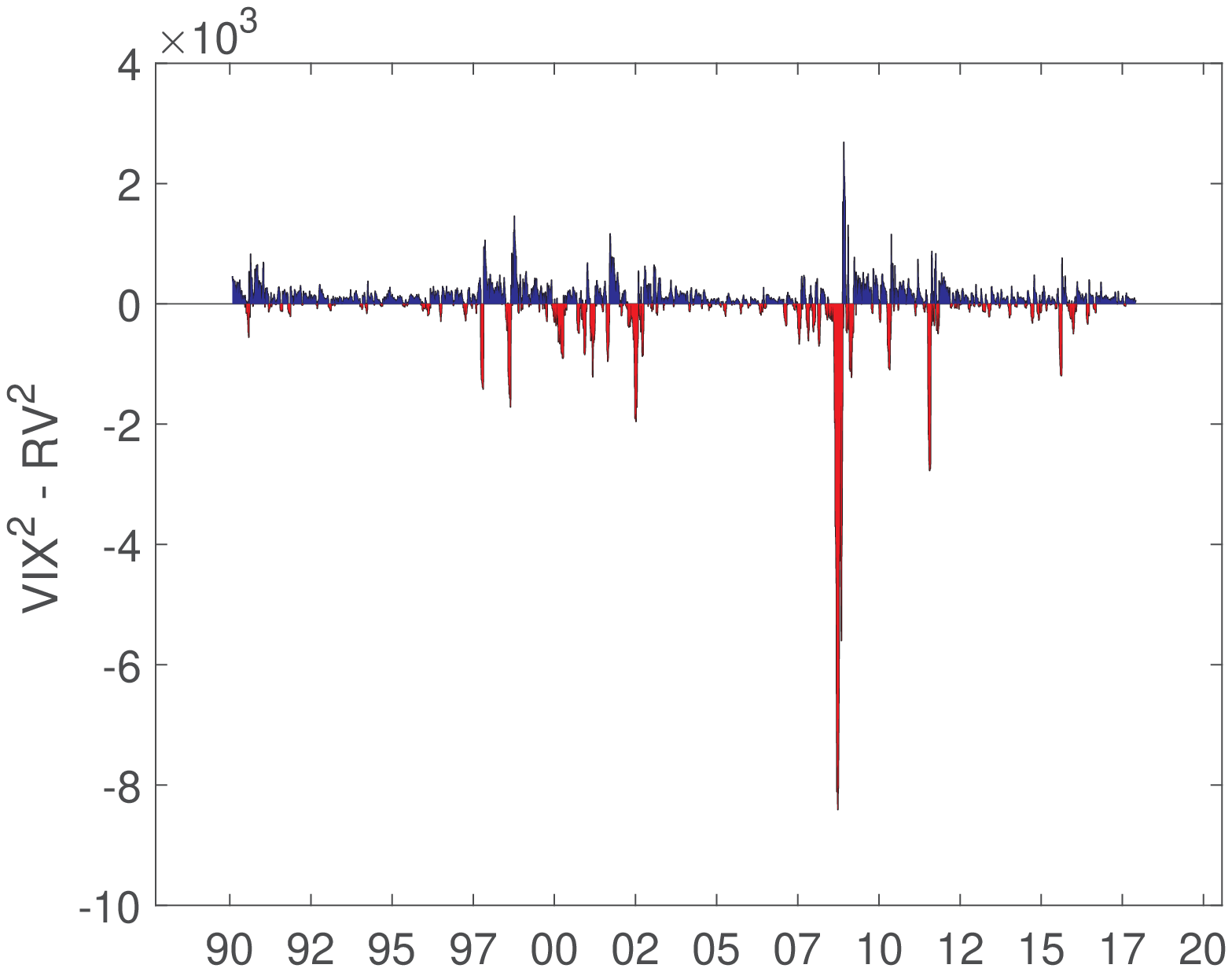}
\includegraphics[width = 0.49 \textwidth]{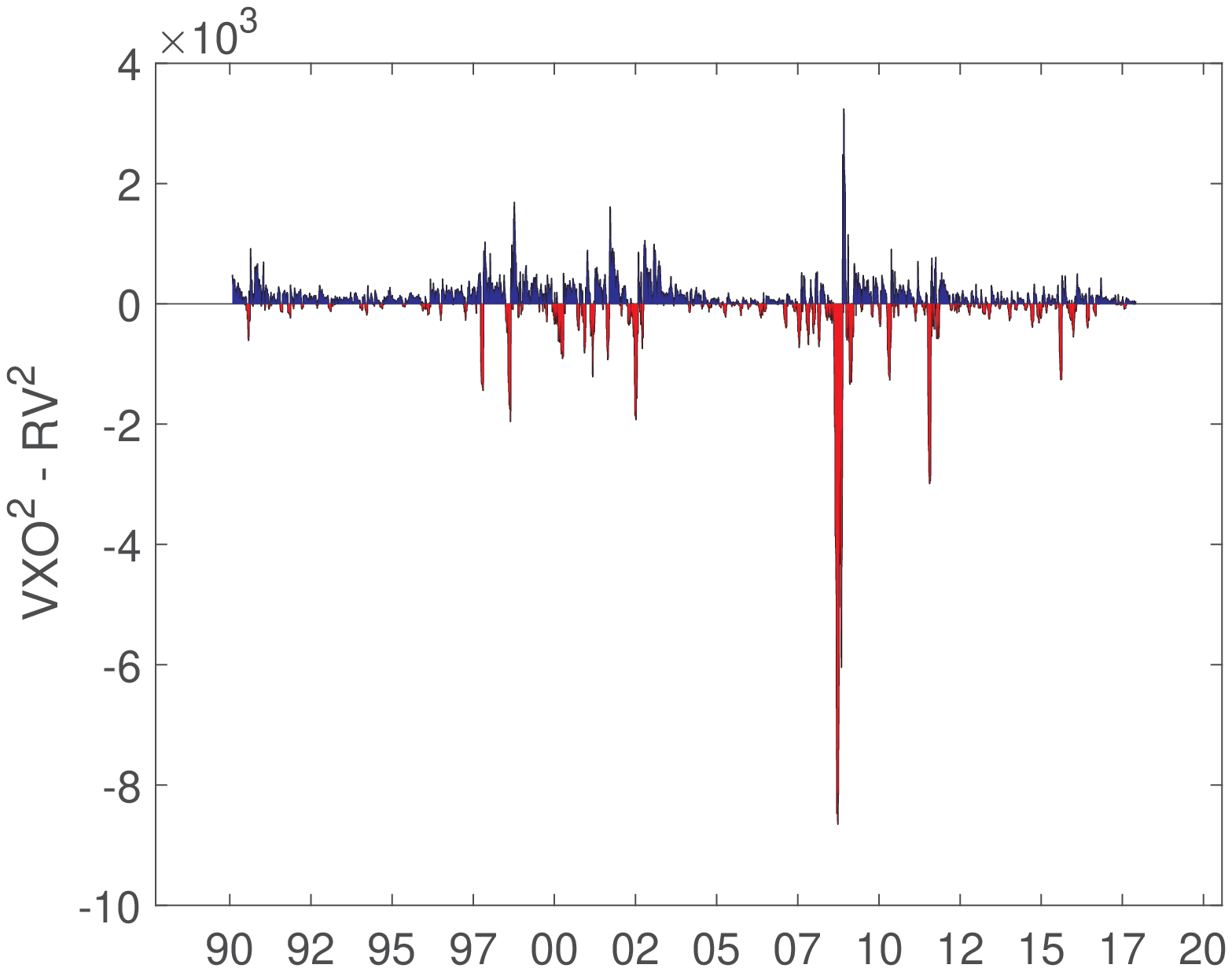}\\
\includegraphics[width = 0.49 \textwidth]{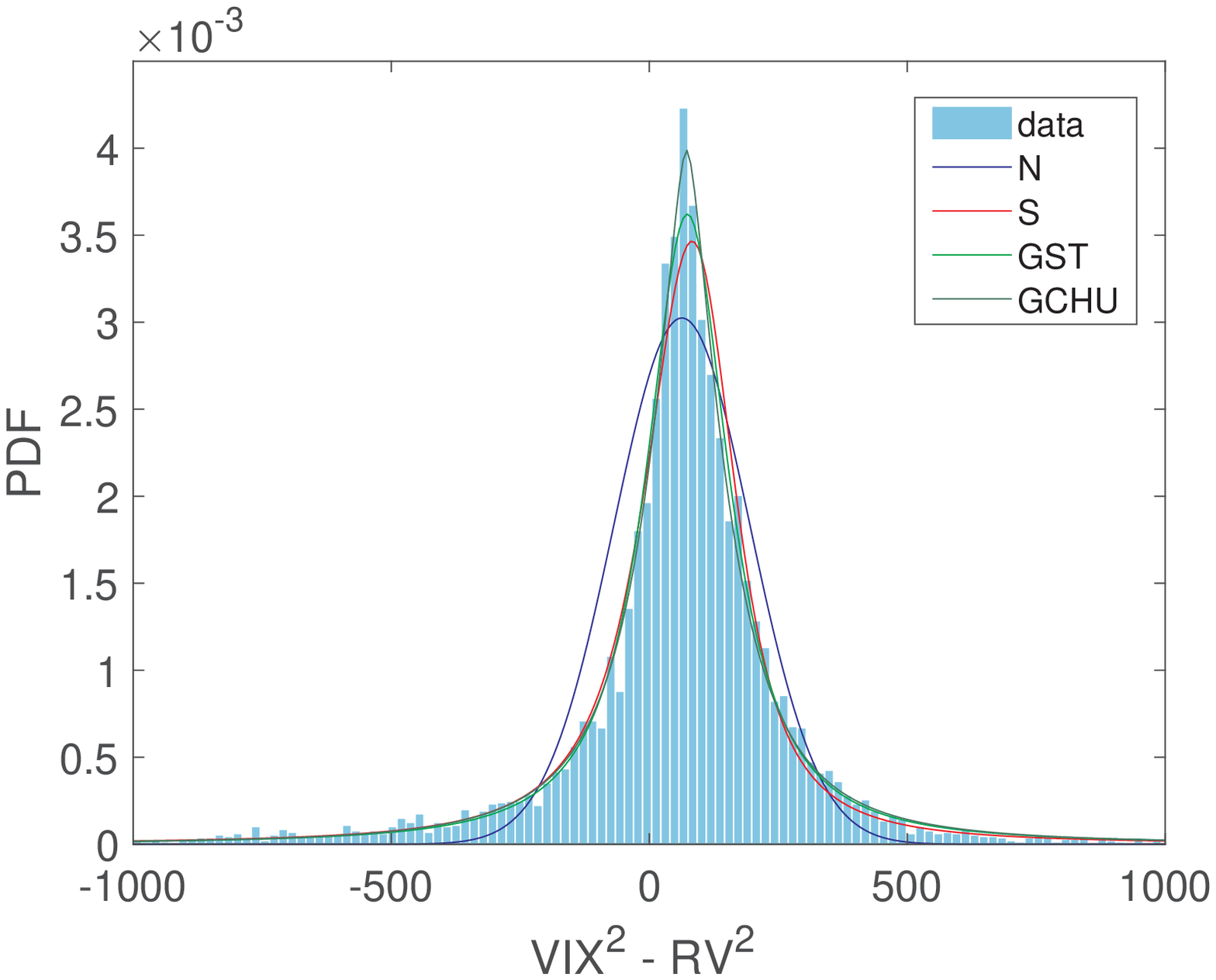}
\includegraphics[width = 0.49 \textwidth]{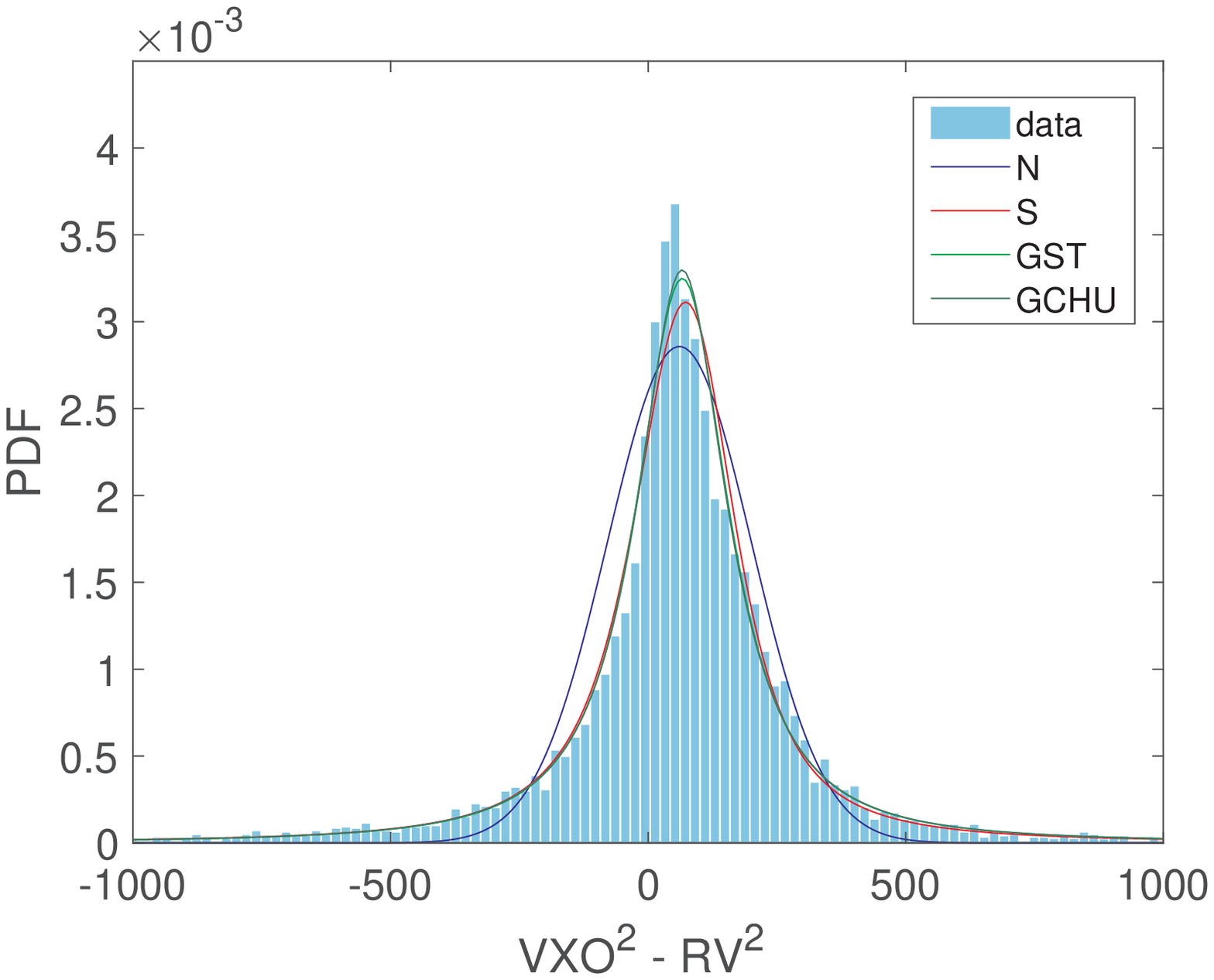}
\end{tabular}
\caption{PDF of $VIX^2 - RV^2$ (left) and $VXO^2 - RV^2$ from Jan 31st, 1990 to Dec 29th, 2017.}
\label{histogramVXO-RV29017}
\end{figure}

In addition to the stable distribution, $S(x; \alpha, \beta, \gamma, \delta)$, discussed in Section \ref{distribution}, we use the three distributions listed in Table \ref{analyticformsdiff}: Normal (N) and two fat-tailed distributions --  Generalized Student's $t$ (GST) and the distribution generalized from the Tricomi Confluent Hypergeometric (GCHU) \cite{dashti2018combined}; the latter, to the best of our knowledge, has not been previously used in the literature for fitting purposes. For these functions, $\mu$ is a location parameter, $\sigma$ is a scale parameter and $\nu$, $p$ and $q$ are shape parameters. The results of fitting are shown in Fig. \ref{histogramVXO-RV29017} and the parameters of the distributions and KS statistics derived from MLE fitting are collected in Tables \ref{MLEVIXminusRV29017} and \ref{MLEVXOminusRV29017}. Notice, that the location parameters for all are rather close to each other and that S, GST, and GCHU KS numbers are very close. In contrast, in Appendix we find that S fit is far more accurate than GST and GCHU for VIX-RV and VXO-RV.

\begin{table}[!htbp]
\centering
\caption{Rescale Values for $RV^2$ for $VIX^2 - RV^2$ (left) and $VXO^2 - RV^2$ (right)}
\label{ratios}
\fontsize{15}{20}
\begin{tabular}{c c}
\hline
	$\frac{mean({VIX}^{2})}{mean({RV}^{2})}=1.4075$&		$\frac{mean({VXO}^{2})}{mean({RV}^{2})}=1.4908$\\
\hline
\end{tabular}
\end{table}

\begin{table}
\centering
\caption{Analytic Form of Distributions For Fitting $VIX^2 - RV^2$ and $VXO^2 - RV^2$}
\label{analyticformsdiff}
\fontsize{15}{20}
\begin{tabular}{|c|c|} 
\hline
            type &       PDF  \\
\hline
$N(x; \mu, \sigma)$ & $\frac{e^{-\frac{(x-\mu )^2}{2 \sigma ^2}}}{\sqrt{2 \pi } \sigma }$ \\
\hline
$GST(x; \mu, \sigma, \nu)$ & $\frac{\left(\frac{\nu }{\nu +\frac{(x-\mu )^2}{\sigma ^2}}\right)^{\frac{\nu +1}{2}}}{\sqrt{\nu } \sigma  B\left(\frac{\nu }{2},\frac{1}{2}\right)} $\\
\hline
$GCHU(x; p, q,  \sigma, \mu)$ & $\frac{\Gamma \left(q+\frac{1}{2}\right) U\left(q+\frac{1}{2},\frac{3}{2}-p,\frac{(x-\mu)^2}{2  \sigma ^2}\right)}{\sqrt{2 \pi } \sigma \Beta(p,q)}$\\
\hline
\end{tabular}
\begin{tablenotes}
\fontsize{10}{15}
\centering
  \item[*]
  \end{tablenotes}
   \end{table}

\begin{table}[!htbp]
\centering
\caption{MLE results for $VIX^2 - RV^2$}
\label{MLEVIXminusRV29017}
\begin{tabular}{ccc} 
\hline
            type &       parameters &          KS test  \\
\hline
Normal & N(         63.2773,         131.8926) &           0.0751 \\
\hline
Gen-Student's $t$ & GST(         73.8714,          92.4056,           1.3310) &           0.0280 \\
\hline
Tricomi & GCHU(          1.7775,           0.7367,          71.2039,          72.3703) &           0.0262 \\
\hline
Stable & S(          1.1842,          -0.1503,          86.5044,          77.5295) &           0.0265 \\
\hline
\end{tabular}
\end{table}

\begin{table}[!htbp]
\centering
\caption{MLE results for $VXO^2 - RV^2$}
\label{MLEVXOminusRV29017}
\begin{tabular}{ccc} 
\hline
            type &       parameters &          KS test  \\
\hline
Normal & N(         60.6005,         139.6113) &           0.0607 \\
\hline
Gen-Student's $t$ & GST(         66.1158,         103.6974,           1.3909) &           0.0245 \\
\hline
Tricomi & GCHU(          8.3382,           0.7080,          31.2797,          65.7904) &           0.0236 \\
\hline
Stable & S(          1.2111,          -0.0899,          95.8820,          69.2693) &           0.0248 \\
\hline
\end{tabular}
\end{table}

\section{Conclusions \label{conclusions}}

We set out to analyze the distribution function of realized variance. We found that it saturates rapidly to its final form after several days of adding daily realized variances. This saturation is quite remarkable in that the daily distribution, with a maximum at low variance and a fat tail with the exponent around 2, gives way to a bell-shaped distribution with strongly suppressed low variance and a fat tail with exponent around 3. The only explanation we can offer is the rapid initial drop-off of correlations of daily variances. We also found that for any number of added days, Generalized Beta Prime distribution would give the best fit. However, all the fitting distributions did poorly for daily realized variance, which is not bell-shaped and is possibly better described by jump models. 

For monthly distributions, we found that squared VIX and VXO distributions are fitted considerably less accurately than the distribution of realized variance. This may be assign of misalignment between implied and realized variance. We also observe a noticeable difference between the 1970-2017 distribution and its 1990-2017 subset, which may indicate that the introduction of implied volatility index influenced future realized volatility. The front end (low volatilities) of all studied distributions is strongly suppressed.

We analyzed the dependence of the fitting parameters of the distribution of realized variance on the number of days over which the daily realized variances are added. We find that the Generalized Beta Prime achieved the fat-tail exponent saturation over about the same number of days as the saturation of the whole distribution. We also find that it very accurately describes the fat tails of the distribution of realized variance. 

Finally, we studied the distribution of difference between squared implied volatility indices and scaled realized variance and found that it is fitted equally well by stable, generalized Student's-$t$ and generalized Tricomi distributions. Among other information, this distribution allows to evaluate the correlation between implied and realized variances.

%\clearpage
\appendix
\section{RV, VIX and VXO Fitting}
\begin{figure}[!htbp]
\centering
\begin{tabular}{cc}
\includegraphics[width = 0.49 \textwidth]{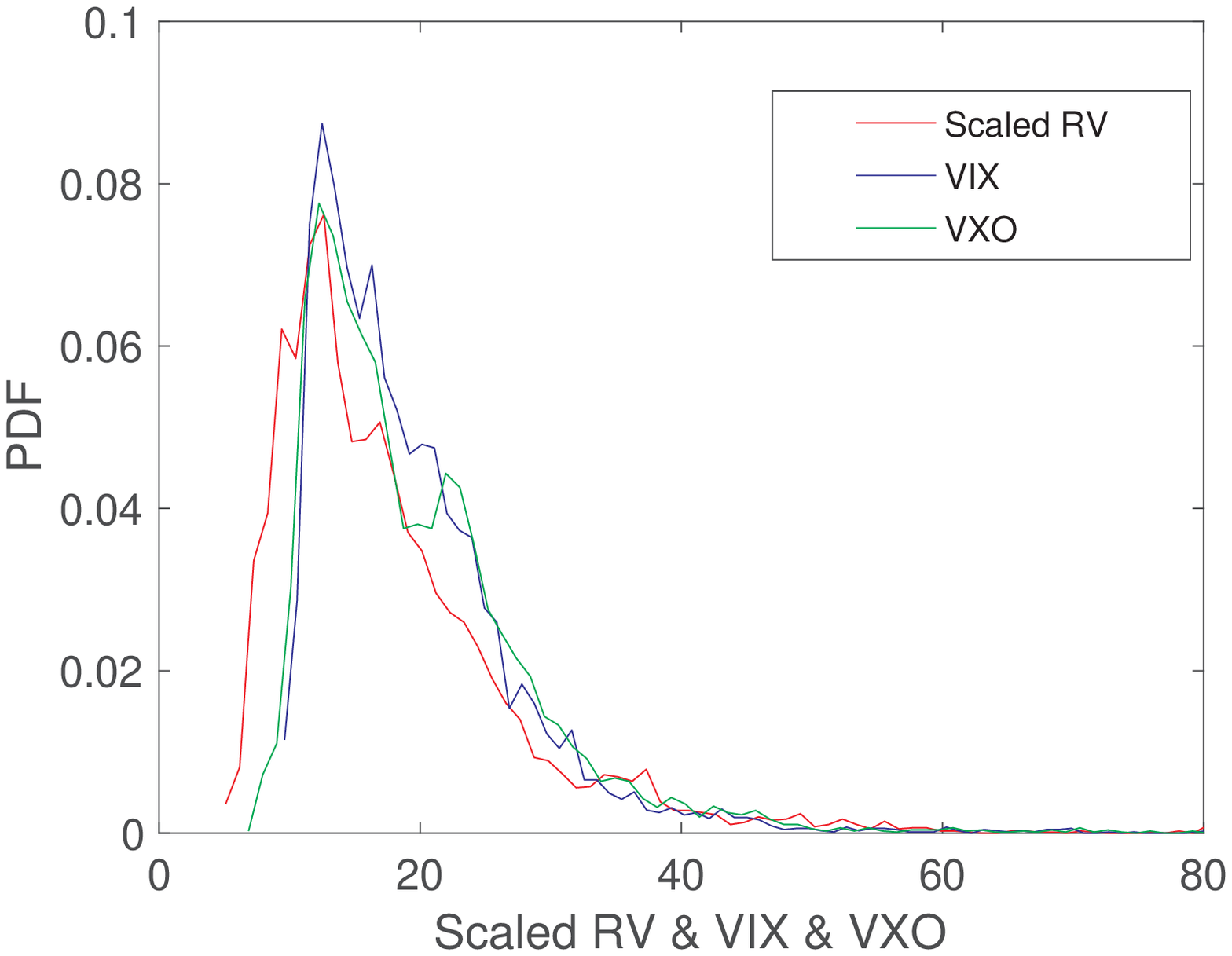}
\includegraphics[width = 0.49 \textwidth]{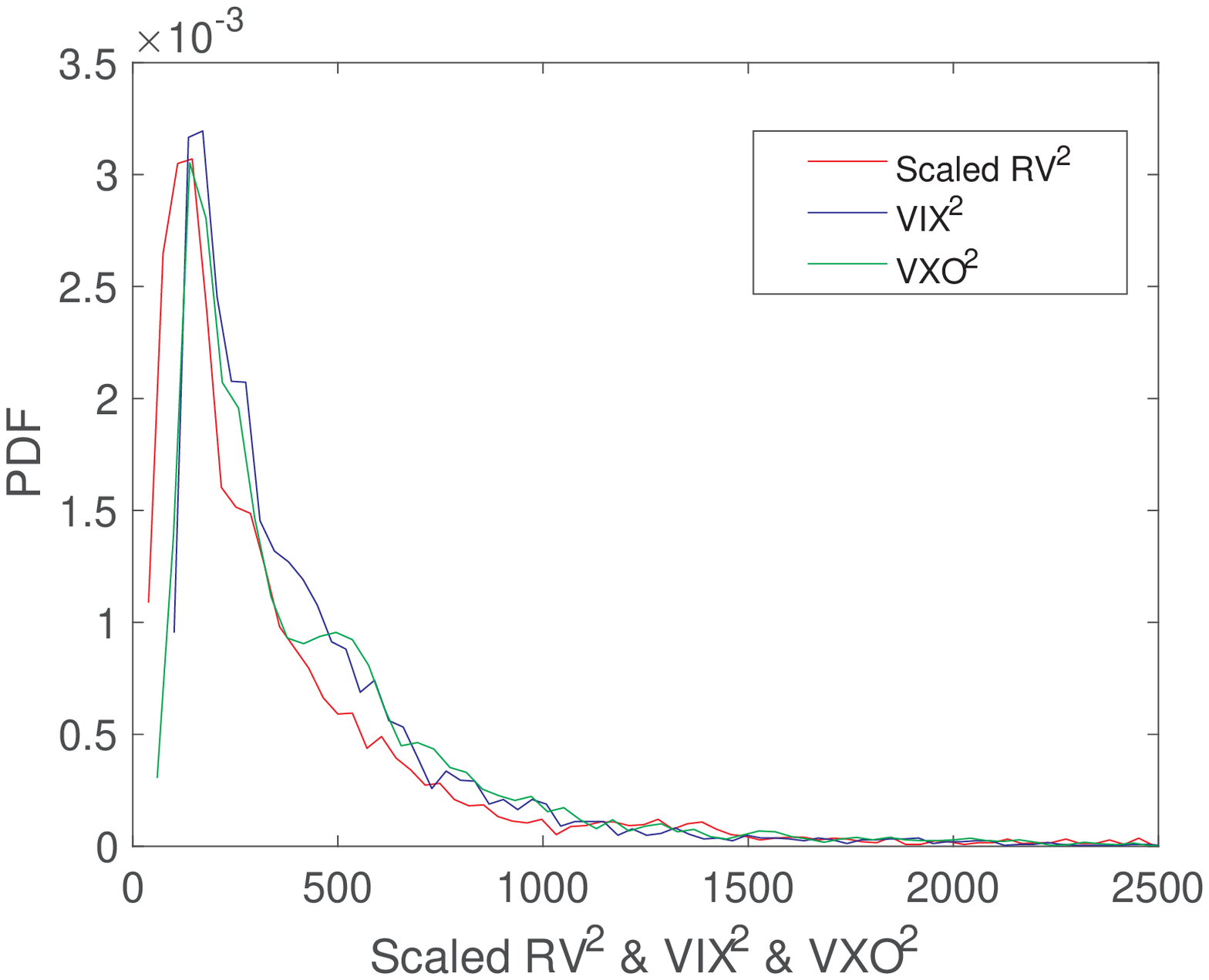}
\end{tabular}
\caption{Contour PDF plots of  scaled RV, VIX and VXO (left) and scaled $RV^2$, $VIX^2$ and $VXO^2$ from Jan 31st, 1990 to Dec 29th, 2017 (right).}
\label{contour}
\end{figure}

\begin{figure}[!htbp]
\centering
\begin{tabular}{c}
\includegraphics[width = 0.49 \textwidth]{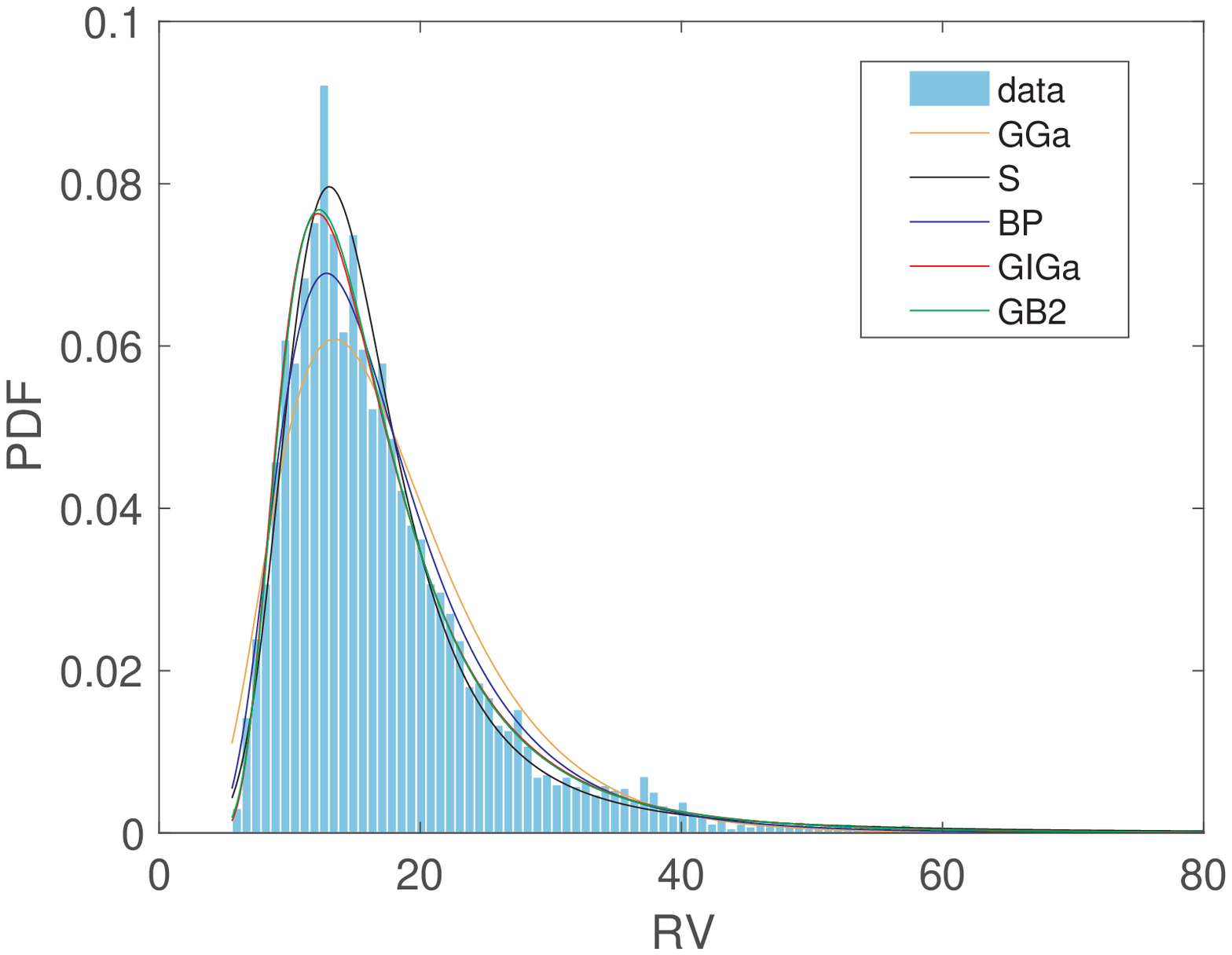}
\includegraphics[width = 0.49 \textwidth]{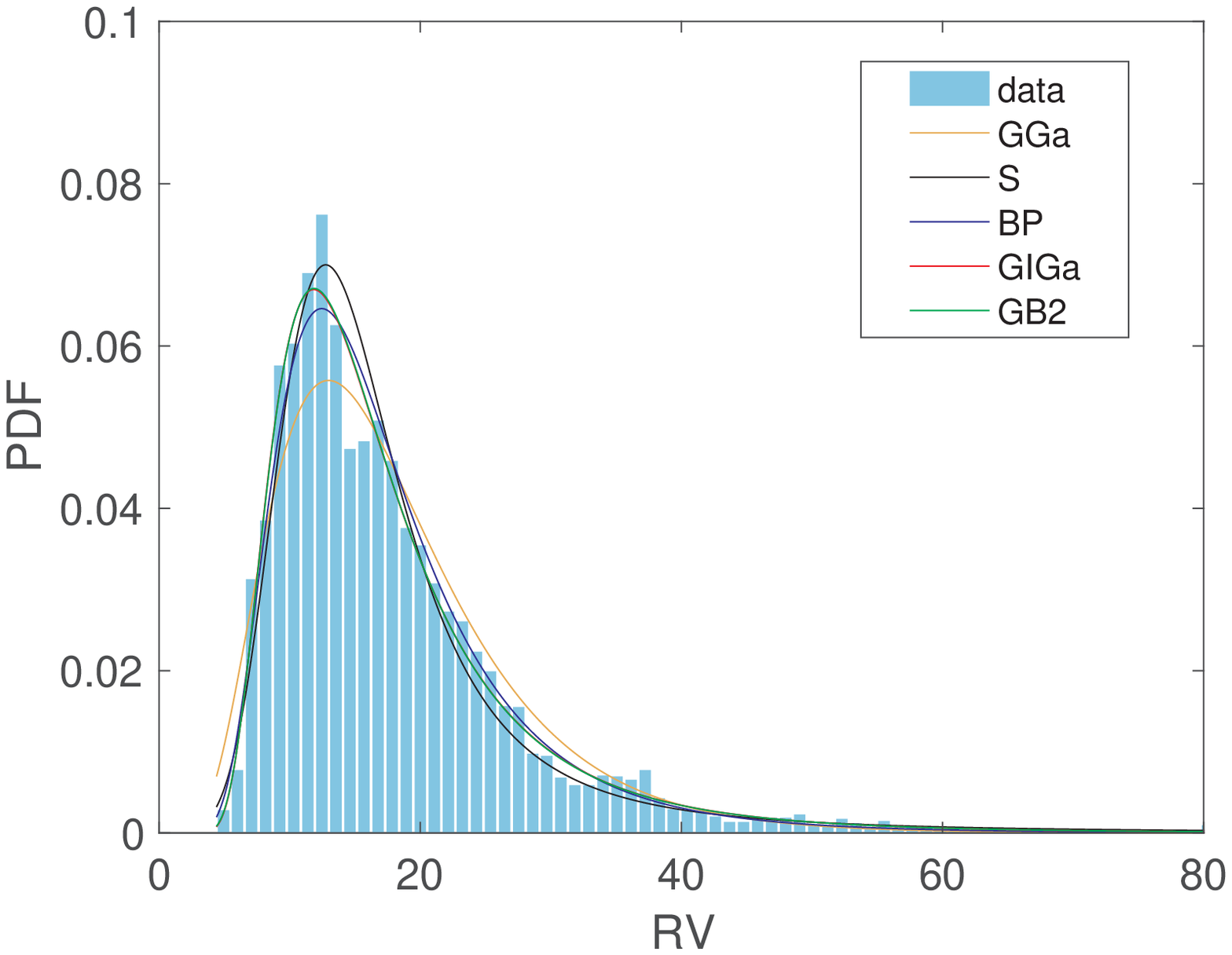}\\
\includegraphics[width = 0.49 \textwidth]{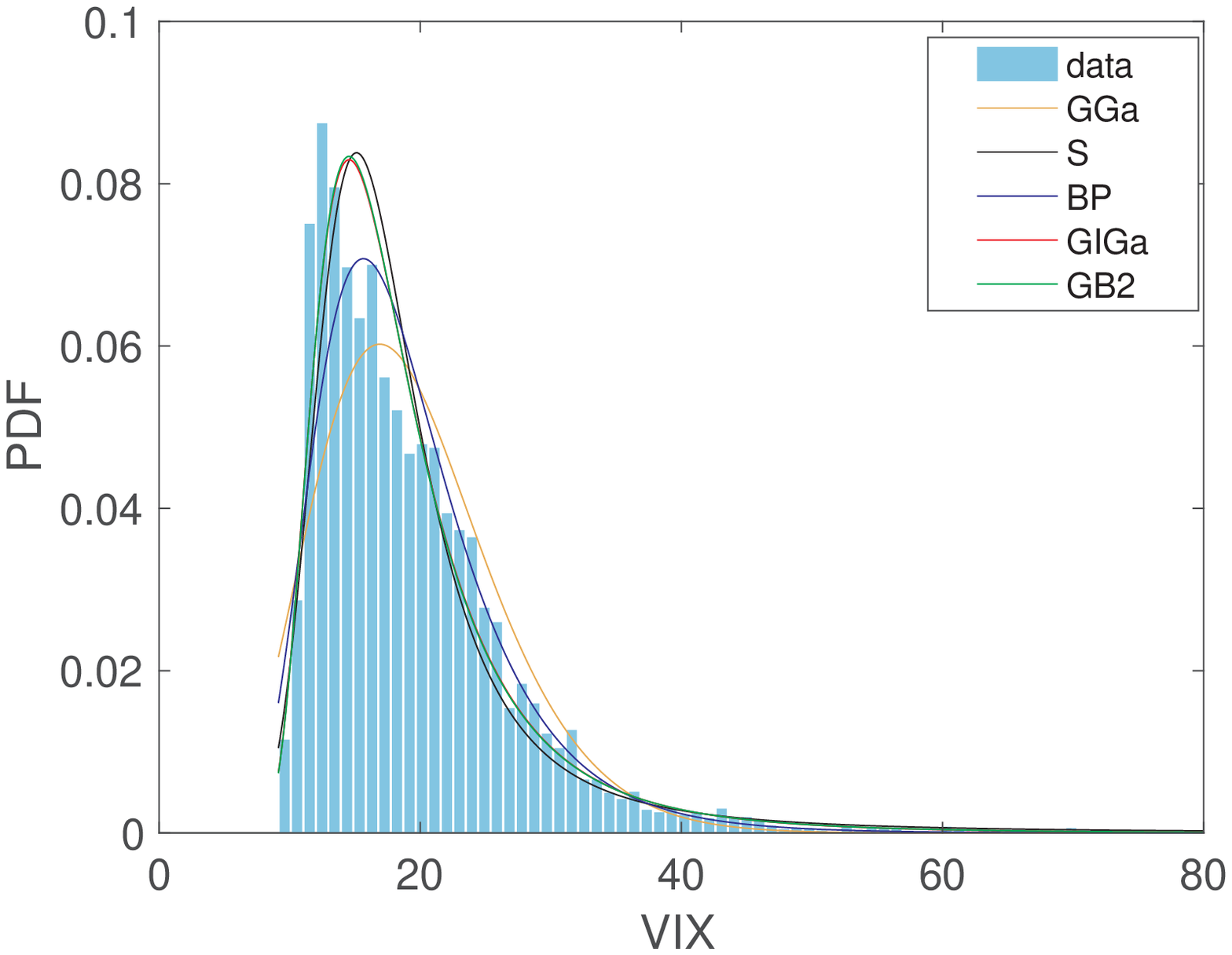}
\includegraphics[width = 0.49 \textwidth]{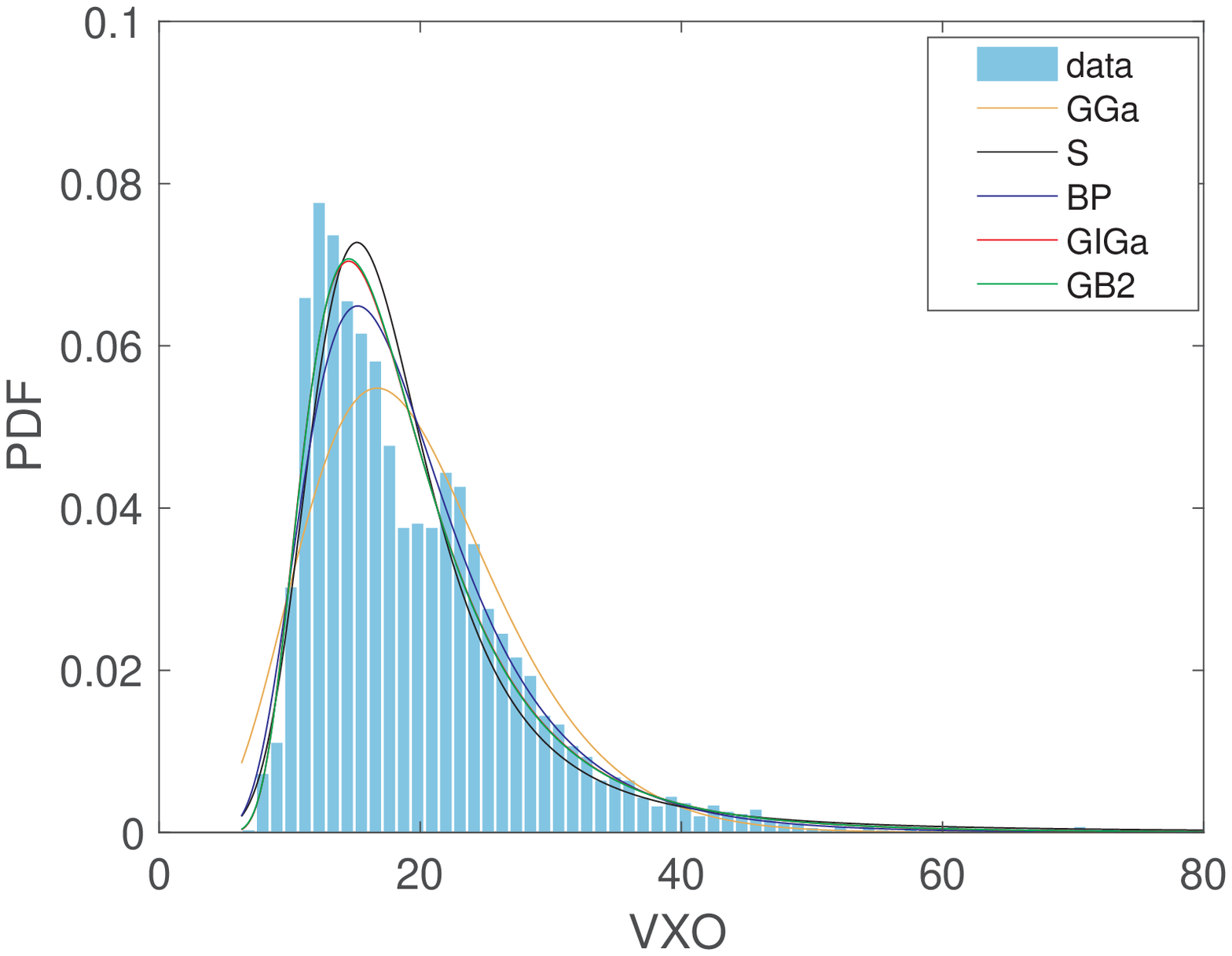}
\end{tabular}
\caption{Clockwise: PDF of monthly RV from Jan 2nd, 1970 to Dec 29th, 2017 and PDFs of monthly RV, VIX and VXO from Jan 31st, 1990 to Dec 29th, 2017.}
\label{indices}
\end{figure}

\begin{figure}[!htbp]
\centering
\begin{tabular}{cc}
\includegraphics[width = 0.49 \textwidth]{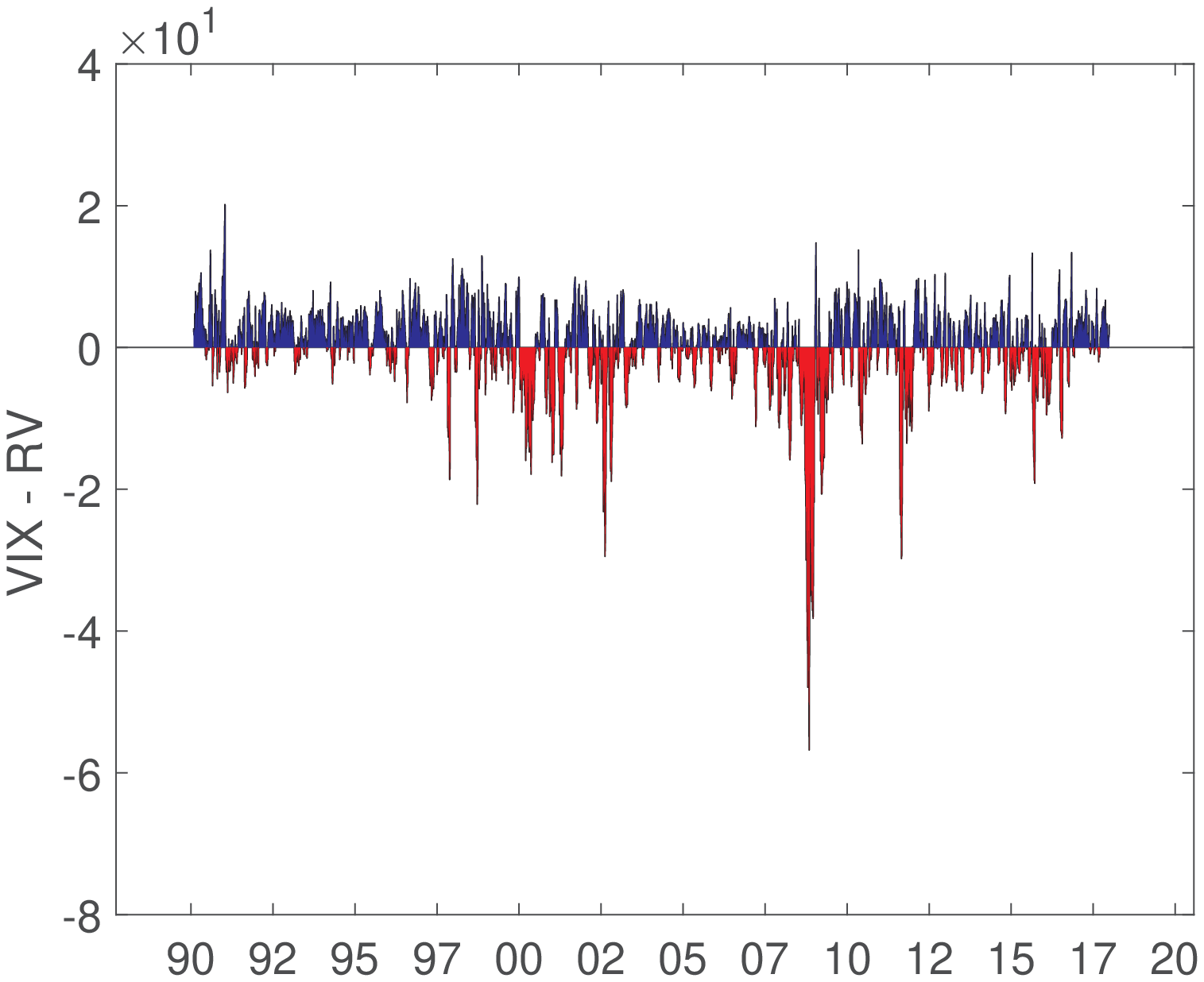}
\includegraphics[width = 0.49 \textwidth]{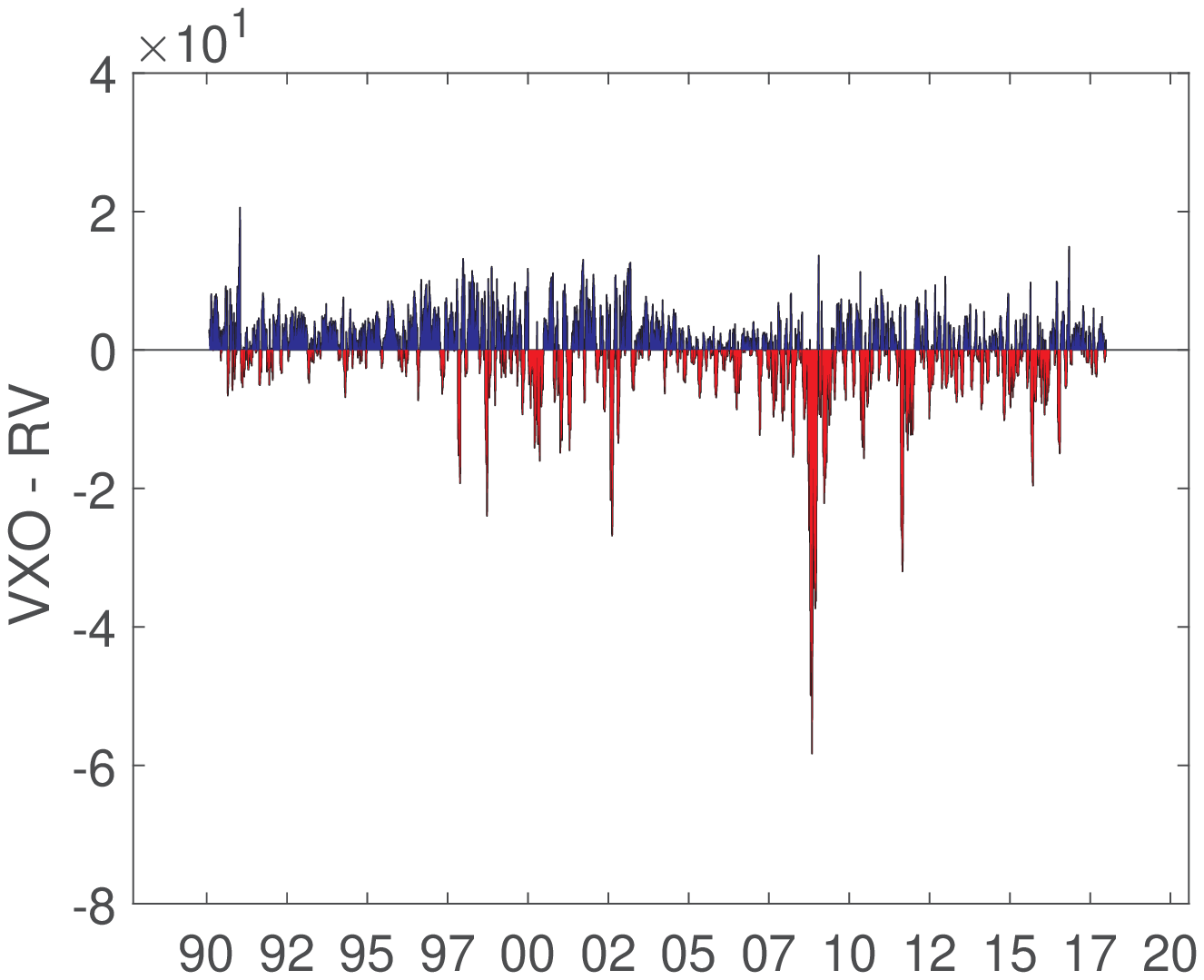}\\
\includegraphics[width = 0.49 \textwidth]{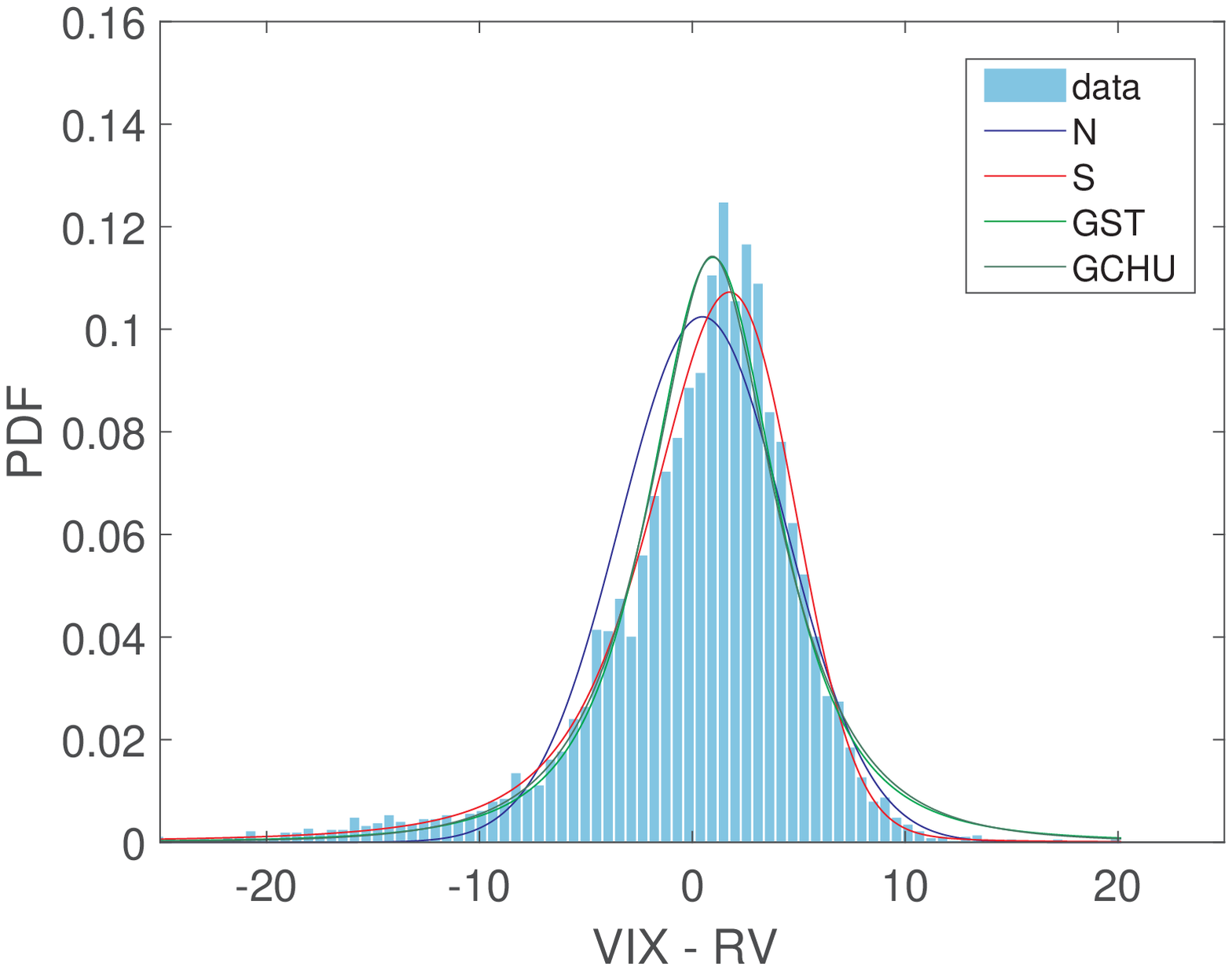}
\includegraphics[width = 0.49 \textwidth]{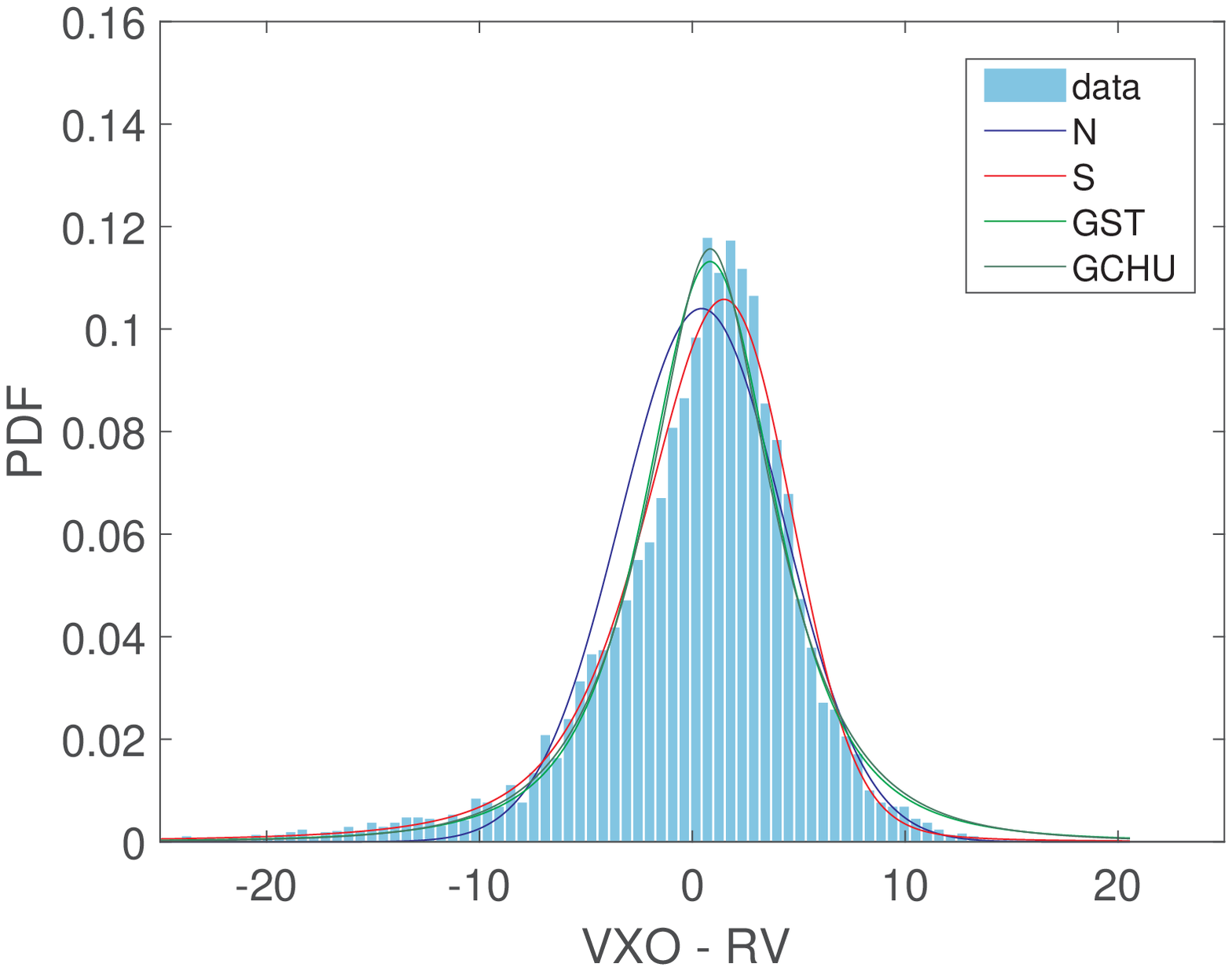}
\end{tabular}
\caption{PDF of $VIX^2 - RV^2$ (left) and $VXO^2 - RV^2$ from Jan 31st, 1990 to Dec 29th, 2017 (right).}
\label{histogramVXO-RV9017}
\end{figure}

Here we fit RV, VIX, VXO and VIX - (scaled) RV and VXO - (scaled) RV. First, for illustrative purposes, in Fig. \ref{contour} we show contour plots of PDF of scaled RV, VIX and VXO vis-a-vis that of scaled $RV^2$, $VIX^2$ and $VXO^2$.
Next, in Fig. \ref{indices} we show fits of the RV, VIX and VXO PDF, with the parameters of the distributions and KS statistics collected in Tables \ref{MLERV7017} - \ref{MLEVXO}. Finally, in Fig. \ref{histogramVXO-RV9017} we show fits of VIX-RV and VXO-RV, with the parameters of the distributions and KS statistics collected in Tables \ref{MLEVIXminusRV9017} - \ref{MLEVXOminusRV9017}.

\clearpage
\begin{table}[!htbp]
\centering
\caption{MLE results for RV 1970-2017}
\label{MLERV7017}
\begin{tabular}{ccccc} 
\hline
            type &       parameters&		front exp&		tail exp&          KS test  \\
\hline
Stable & S(          1.3278,           1.0000,           3.4936,          13.8773) &		 &		 -2.3278&           0.0171 \\
\hline
GB2 & GB2(         15.8782,           1.8724,           2.0309,          4.8757) &		31.2470&		-4.8026&           0.0115 \\
\hline
BP & BP(         27.1723,           6.7415,           4.001) &		26.1723&		-7.7415&           0.0275 \\
\hline
GIGa & GIGa(          2.5562,          25.0090,           1.6047) &		  &		-5.1019 &           0.0138 \\
\hline
IGa & IGa(          6.0553,         88.2509) &		  &		-7.0553 &           0.0164 \\
\hline
GGa & GGa(          6.0715,           2.3493,     0.9052)&		4.4959&		 &           0.0753 \\
\hline
Ga & Ga(          4.8790,           3.6151) &		3.8790&		 &           0.0795 \\
\hline
\end{tabular}
\end{table}

\begin{table}[!htbp]
\centering
\caption{MLE results for RV 1990-2017}
\label{MLERV2}
\begin{tabular}{ccccc} 
\hline
            type &       parameters&		front exp&		tail exp&          KS test  \\
\hline
Stable & S(          1.2849,           1.0000,           3.9402,          13.7767) &		 &		 -2.2849&           0.0249 \\
\hline
GB2 & GB2(         17.6690,           2.4125,           1.5731,          4.1218) &		26.7951&		-4.7951&           0.0139 \\
\hline
BP & BP(         21.8899,           6.1274,           4.2232) &		20.8899&		-7.1274&           0.0298 \\
\hline
GIGa & GIGa(          3.5363,          40.8574,           1.1920) &		 &		-5.2153 &           0.0134 \\
\hline
IGa & IGa(          4.8913,         70.5571)&		  &		-5.8913&           0.0182 \\
\hline
GGa & GGa(          4.3867,           3.9442,     0.9731)&		3.2687&		 &           0.0647 \\
\hline
Ga & Ga(          4.1254,           4.4086) &		3.1254&		 &           0.0659 \\
\hline
\end{tabular}
\end{table}

\begin{table}[!htbp]
\centering
\caption{MLE results for VIX 1990-2017}
\label{MLEVIX}
\begin{tabular}{ccccc} 
\hline
            type &       parameters&		front exp&		tail exp&          KS test  \\
\hline
Stable & S(          1.2901,           1.0000,           3.3182,          15.9886)  &		 &		 -1.2901&           0.0381 \\
\hline
GB2 & GB2(         63.0279,           1.3326,           2.9404,		4.2422)  &		184.3272&		-4.9184&           0.0368 \\
\hline
BP & BP(         44.8997,          10.8471,           4.2232)  &		43.8997&		-11.8471&           0.0463 \\
\hline
GIGa & GIGa(          1.4520,          18.0537,           2.7628) &		  &		-5.0116 &           0.0375 \\
\hline
IGa & IGa(          8.9387,         152.9579) &		  &		-9.9387 &           0.0443 \\
\hline
GGa & GGa(         6.7354,           10.0120,           0.5250) &		2.5360 &		 &           0.0669 \\
\hline
Ga & Ga(          7.7138,           2.5092) &		6.7138 &		 &           0.0681 \\
\hline
\end{tabular}
\end{table}

\begin{table}[!htbp]
\centering
\caption{MLE results for VXO 1990-2017}
\label{MLEVXO}
\begin{tabular}{ccccc} 
\hline
            type &       parameters&		front exp&		tail exp&          KS test  \\
\hline
Stable & S(          1.3267,           1.0000,           3.8227,          16.0853)  &		 &		 -2.3267&           0.0434 \\
\hline
GB2 & GB2(         53.3880,           2.6150,           1.7816,		3.0120)  &		94.1160&		-5.6589&           0.0391 \\
\hline
BP & BP(         36.5647,           8.8725,           4.2232)  &		35.5647&		-9.8725&           0.0436 \\
\hline
GIGa & GIGa(          3.0721,          33.0522,           1.5909) &		  &		-5.8874&           0.0423 \\
\hline
IGa & IGa(          7.2686,         123.1749) &		   &		-8.2686&           0.0491 \\
\hline
GGa & GGa(         5.5107,           3.9062,           1.0601) &		4.8419 &		 &           0.0718 \\
\hline
Ga & Ga(          6.4518,           3.0512)&		5.4518 &		 &           0.0712 \\
\hline
\end{tabular}
\end{table}

\begin{table}[!htbp]
\centering
\caption{MLE results for VIX-RV}
\label{MLEVIXminusRV9017}
\begin{tabular}{ccccc} 
\hline
            type &       parameters &          KS test  \\
\hline
Normal & N(          0.4791,           3.8945) &           0.0673 \\
\hline
Gen-Student's $t$ & GST(          0.9680,           3.1944,           2.6948) &           0.0447 \\
\hline
Tricomi & GCHU(          3.1107,           2.1000,           3.0969,           0.9445) &           0.0467 \\
\hline
Stable & S(          1.5807,          -0.8377,           2.6247,           1.4591) &           0.0135 \\
\hline
\end{tabular}
\end{table}

\begin{table}[!htbp]
\centering
\caption{MLE results for VXO-RV}
\label{MLEVXOminusRV9017}
\begin{tabular}{ccccc} 
\hline
            type &       parameters &          KS test  \\
\hline
Normal & N(          0.4412,           3.8358) &           0.0618 \\
\hline
Gen-Student's $t$ & GST(          0.8367,           3.2302,           2.8099) &           0.0392 \\
\hline
Tricomi & GCHU(          2.8284,           2.1000,           3.2574,           0.8439) &           0.0426 \\
\hline
Stable & S(          1.6068,          -0.7346,           2.6689,           1.2449) &           0.0159 \\
\hline
\end{tabular}
\end{table}

Two peculiarities should be noted. First, in contrast to $VIX^2-RV^2$ and $VXO^2-RV^2$, where S, GST and GCHU fitted equally well, here we find that S fit is more accurate for VIX-RV and VXO-RV, probably because of a greater skewness of the latter two. Second, BP fits are worse than IGa for RV but are  better for $RV^2$. Obviously, distributions of RV and $RV^2$ are not independent: under transformation $x \rightarrow x^r, r > 0$ we have $GB2(x; p, q, \alpha, \beta)\rightarrow GB2(x; p, q, \alpha r, \beta^{1/r})$ and $GIGa(x; \alpha, \gamma, \beta) \rightarrow GIGa(x; \alpha, \gamma r, \beta^{1/r})$. With the values of parameters for RV, IGa transforms into GIGa which fits $RV^2$ distribution rather well, if not as the best GIGa fit, while neither BP nor GB2 with $RV^2$ parameters transforms into a GB2 that is close to BP.

\clearpage
\bibliography{mybib}

\end{document}